\documentclass[a4paper,fleqn,usenatbib]{mnras}

\usepackage{newtxtext,newtxmath}

\usepackage[T1]{fontenc}
\usepackage{ae,aecompl}

\usepackage{bm}

\usepackage{graphicx}	
\usepackage{amsmath}	
\usepackage{amssymb}	
\usepackage{color}

\newcommand{\NHVS}{{N_{\text{HVS}}}}
\newcommand{\bw}{\mathbfit{w}}
\newcommand{\bx}{\mathbfit{x}}
\newcommand{\bv}{\mathbfit{v}}
\newcommand{\Gaia}{{\it Gaia} }

\title[On measuring the Galactic halo with HVSs]{On measuring the Galactic dark matter halo with hypervelocity stars}

\author[O.~Contigiani et al.]{
O.~Contigiani,\thanks{E-mail: contigiani@strw.leidenuniv.nl}
E.~M.~Rossi,
T.~Marchetti
\\
Leiden Observatory, Leiden University, PO box 9513, NL-2300 RA, Leiden, the Netherlands\\
}

\date{Accepted XXX. Received YYY; in original form ZZZ}

\pubyear{2018}

\begin{document}
\label{firstpage}
\pagerange{\pageref{firstpage}--\pageref{lastpage}}
\maketitle

\begin{abstract}
{ Hypervelocity stars (HVSs) travel from the Galactic Centre across the dark matter halo of the Milky Way, where they are observed with velocities in excess of the Galactic escape speed. Because of their quasi-radial trajectories, they represent a unique probe of the still poorly constrained dark matter component of the Galactic potential. In this paper, we present a new method to produce such constraints. Our likelihood is based on the local HVS density obtained by back-propagating the observed phase space position and quantifies the ejection probability along the orbit. To showcase our method, we apply it to simulated \Gaia samples of $\sim200$ stars in three realistic Galactic potentials with dark matter components parametrized by spheroidal NFW profiles. We find that individual HVSs exhibit a degeneracy in the scale mass-scale radius plane ($M_s-r_s$) and are able to measure only the combination $\alpha = M_s/r_s^2$. Likewise, a degeneracy is also present between $\alpha$ and the spheroidal axis-ratio $q$. In the absence of observational errors, we show the whole sample can nail down both parameters with {\it sub-percent} precision (about $1\%$ and $0.1\%$ for $\alpha$ and $q$ respectively) with no systematic bias. This remarkable power to constrain deviations from a symmetric halo is a consequence of the Galactocentric origin of HVSs. To compare our results with other probes, we break the degeneracy in the scale parameters and impose a mass-concentration relation. The result is a competitive precision on the virial mass $M_{200}$ of about $10\%$. }
\end{abstract}




\section{Introduction}
In the concordance $\Lambda$CDM model of cosmology, galaxies are embedded inside larger structures known as haloes. These are made of a dissipationless fluid called dark matter, visible only through its gravitational effects. Over cosmic time, haloes grow in mass and size through hierarchical clustering, starting from the initial perturbations of a slightly inhomogeneous matter density field. Despite its central role in structure formation, the nature of dark matter and its microscopic physics are still unknown \citep[see, e.g.,][]{Garrett2011}. 

There is a number of theoretical predictions associated to the shape and mass of dark matter structures. Pure cold dark matter simulations suggest that collapsed haloes acquire a triaxial ellipsoid shape, but more recently it has been found that the inclusion of baryonic matter results in rounder shapes \citep[e.g.,][]{Debattista2008}. Similarly, self interacting dark matter is also expected to induce spherical haloes in the innermost regions \citep{Peter2013}. In this context, measurements of the Milky Way's halo, together with observations of surrounding dwarf galaxies, can be used as a test for the concordance model \citep{Moore1999, Klypin1999}. For example, a total mass of the Milky Way lower than $10^{12}$ $M_{\odot}$ can align the observed number of satellite galaxies with what is predicted in simulations \citep{Wang2011}. 

Gravitational lensing is the most common technique used to measure the dark matter distributions of statistical samples of distant galaxies and galaxy groups \citep[e.g.,][]{Hoekstra2013, Mandelbaum2014}. In the case of the Milky Way, our privileged position mandates the use of a different set of techniques and dynamical tracers are employed to measure the structure of its dark matter halo. Objects travelling through the halo act as test particles subjected to its gravitational potential and their trajectories in phase space can be traced to constrain a parametric model for the density profile. This procedure usually requires assumptions about the initial conditions or the steady-state configuration of the system.

In the Galactic bulge and disc, where baryons dominate the matter density, established techniques based on the kinematics of field stars or HI emission are used  \citep{Portail2017, Reid2016}. Unfortunately, the scarcity of these tracers outside the Galactic disc limits their constraining power where the dark matter halo dominates \citep[e.g., ][]{Huang2016}. 
Since its discovery \citep{Newberg2002}, the Sagittarius stellar stream has proved to be a valuable dynamical tracer in this region \citep[e.g.,][]{Law2009, Deg2013, Gibbons2014}. The Sagittarius dwarf galaxy is one of the closest satellites to the Milky Way and it is in the process of being tidally disrupted. The strong tidal forces give rise to a long stream of tidal debris which orbits the Milky Way. Other tidal streams have been discovered over the years: some of them connected to globular clusters \citep[e.g. Palomar 5,][]{Odenkirchen2001} and some others represent the last remnants of now defunct dwarf galaxies \citep[e.g. Virgo,][]{Duffau2014}.

Despite the existence of these multiple tracers, there is no consensus in the literature on the mass of the Milky Way halo  \citep{Wang2015}: measurements differ up to a factor $5$ and relative precisions range from below $10\%$ to roughly $100\%$. The situation is no different when, instead of its mass, the halo shape is considered. While the Milky Way's dark matter halo is often measured to be a spheroid with two of its axes being equal and aligned with the disc galaxy within \citep[e.g.,][]{Bovy2015, Pearson2015}, conflicting measurements are present in the literature and triaxial shapes have also been suggested \citep[e.g.][]{Law2010}. The halo shape could also be a function of radius, spheroidal in the centre and triaxial in the outer region  \citep{Vera-Ciro2013}. In the case of a pure spheroid, the ratio between the third axis and one of the others is usually referred to as $c/a$ or, like in this paper, just $q$. A ratio $q=1$ corresponds to a sphere, while the conditions $q>1$ and $q<1$ correspond respectively to a prolate or an oblate spheroid. Reports range from a spherical halo \citep[e.g.,][]{Bovy2016} to oblate \citep[e.g.,][]{Loebman2014} or prolate \citep[e.g.,][]{Bowden2016, Posti2018}.  It is clear that when previous endeavours to measure the Galactic halo are put together, the tensions between different probes imply the existence of systematic biases. 

In future years, hypervelocity stars (HVSs) are expected to be introduced to this landscape as a powerful probe. For the purposes of this work we will refer to HVSs as high velocity objects (Galactocentric velocity $>450$ km/s) travelling from the Galactic Centre (GC) along quasi-radial orbits. In 2005 the first HVS was discovered  \citep{Brown2005}: a B-type main-sequence star with radial velocity in the Galactic rest frame of about $700$ km/s. Subsequent observations have measured its distance from the GC, found to be of the order of $100 $ kpc \citep{Brown2014}. Given its high velocity, the object was measured to be unbound form the Galaxy. Over the years, objects with similar stellar properties have been found and to date the largest and most studied sample is composed of the 21 HVS candidates reported by \cite{Brown2014}, a survey targeting B-type stars in the outer halo. In the near future, the high-quality sample of HVSs predicted to be observed by the satellite mission \Gaia \citep{GaiaCollaboration2016, GaiaCollaboration2018} by the early 2020s is expected to contain several hundred objects \citep{Marchetti2018} and will offer a new diffuse dynamical tracer for the Galactic potential. 

The goal of this paper is to introduce a new method to exploit this tracer. \cite{Gnedin2005} already showed that a few HVSs can be a powerful tool to constrain the shape and orientation of the Galactic halo and a precision of about $10\%$ can be reached if accurate proper motion and Galactocentric distances are known. Later, \cite{Yu2007} have shown how the triaxiality of an ellipsoidal halo can be estimated directly from observed HVS positions and velocities under a specific halo model. Other similar attempts include \cite{Perets2009}, who explored how asymmetries in the radial velocity distribution of halo stars due to HVSs depend on the Milky Way mass, and \cite{Fragione2017}, which is an application of such method. In other cases, inferences about the Galactic gravitational potential behind the deceleration of HVSs assume a certain class of ejection velocity distributions \citep[e.g.,][]{Sesana2007, Rossi2016}.

We expand on previous works by developing a new versatile technique that can be adapted with minimal assumptions to a variety of models for the ejection mechanism and Galactic potential. This is of the uttermost importance to produce unbiased joint constraints in combination with other probes \citep[see][where two of us have shown the power of this approach]{Rossi2016}. Our method is based on a reconstruction of the HVS orbital history and it has the advantage of not requiring simulations of the entire population for every potential/ejection model studied.\footnote{In the interest of reproducibility we make our code publicly available at \url{https://www.github.com/contigiani/hvs}.}

To validate our method we will focus here on HVSs ejected through one realization of the Hills mechanism \citep{Hills1988, Yu2003, Sari2010}. According to this mechanism, the three-body interaction between a binary system and a massive black hole (MBH) results in one star orbiting closely around the black hole and the other one being ejected at high velocity. The aforementioned observations of high velocity stars in the Galactic halo are consistent with the existence of such mechanism and, at present, it is still considered the leading explanation \citep{Brown2015, Brown2018}. Note also that HVSs are expected to be an observational consequence of the massive black hole located in the GC \citep[][]{Ghez2003}.

In Sec. \ref{sec:numsim} we construct mock populations of this sample, based on previous work \citep{Rossi2013, Rossi2016, Marchetti2018}. Our mock catalogues are based on the expected astrometry and photometry of the final \Gaia data release. Afterwards, we lay the foundations of our technique and we arrive in Section \ref{sec:methods} at an integral formula for the phase space distribution of these objects, which allows us to write down a likelihood function for an observed sample of HVSs. In the same section, we also discuss the advantages and limitations of the method. In Sec. \ref{sec:res} we then test our approach and try to recover the dark matter halo inside which the simulated sample was propagated. In the same section, we also address issues related to the practical implementation of our technique for a \Gaia-like sample.

\section{Simulated HVS catalogues}
\label{sec:numsim}
The first step to verify how and if HVSs can constrain the dark matter halo of the Milky Way is to produce an observational mock catalogue of HVSs. 
To produce such sample we need to specify three important ingredients: 1) the ejection distribution that determines how the velocities, positions and masses of our stars are distributed at the moment of ejection from the GC; 2) a survival function that dictates the fraction of HVSs alive after a time $t$ post-ejection, and 3) a gravitational potential under the influence of which the stars trace their orbits. In the next three subsections we present an implementation of these quantities and we conclude, in the last subsection, with the details of our numerical simulation. 

\subsection{Ejection rate distribution}

We aim to parametrize the distribution of velocities, positions and masses at ejection for HVSs generated through the Hills mechanism \citep{Hills1988} by writing down an explicit expression for an ejection rate distribution $\mathcal{R}(\bw)$, which has the units of a configuration space density per unit time. We call $\bw$ our configuration space coordinate, $\bw = (\mathbfit{x}, \mathbfit{v}, m)$, where $(\mathbfit{x}, \mathbfit{v})$ is the usual phase space coordinate (position, velocity) and $m$ is the stellar mass. We follow the set up first described in \cite{Rossi2016} and then implemented by \cite{Marchetti2018}. 

In a reference system centred on the massive black hole (or equivalently, the GC) we can write:
\begin{equation}
\mathcal{R}(\bw=(\bx, \bv, m)) = \Gamma \, \mathcal{R}_{H}(|\bv|, m) \, \delta \left( |\bx| \right) \, \delta \left( \bx\cdot\bv \right),
\label{eq:R}
\end{equation}
where we have introduced the ejection rate per unit time $\Gamma$ and the $\delta$ terms are Dirac deltas. In this work we will not assume any value for $\Gamma$ and we will normalize all of the other functions appearing in this expression to unity. 

The main prediction of the Hills mechanism quantifies the asymptotic velocity of the ejected objects at an infinite distance from the massive black hole. In practice, this distance can be modelled as the radius of the gravitational sphere of influence of the black hole $\bar{r}$, defined as the radius of the sphere centred on the black hole and containing twice its mass. For distances larger than its radius the potential of the black hole becomes a negligible fraction of the total Galactic potential.  We pick the value $\bar{r} = 3$ pc \citep{Genzel2010} and impose this to be the ejection radius through the Dirac delta function $ \delta \left( |\bx| \right)$.

The term $ \mathcal{R}_{H}(|\bv|, m)$ quantifies the relative probabilities of different initial velocities and masses of HVSs. It can be computed using Monte Carlo (MC) simulations as done in \cite{Rossi2013, Rossi2016, Marchetti2018}. In the first paper it is also shown that the resulting distributions can be easily fitted with analytic functions. By fitting the Hills mechanism MC catalogue in \cite{Marchetti2018} to the functional form suggested by \cite{Rossi2013} we obtain

\begin{align}
\begin{cases}
\mathcal{R}_{H}(|\bv|, m) \propto m^{-1.7} |\bv|^{-1}
&\text{if } |\bv| \leq v_0(m),
\\
\mathcal{R}_{H}(|\bv|, m) \propto m^{-1.7} |\bv|^{-6.3} & \text{if } |\bv| > v_0(m);
\end{cases}
\label{eq:Rvm}
\end{align}

\begin{align}
v_0(m) = 1530~ (\text{M}_\odot/m)^{0.65} \text{ km/s}.
\end{align}

Notice that the velocity distribution for a fixed value of $m$ has a high velocity tail starting from the value $v_0(m)$. 

The last term in  eq.~\ref{eq:R} is a Dirac delta function imposing zero angular momentum. This condition must be satisfied at any ejection distance $|\bx|>\bar{r}$ since every HVS is a product of a close encounter of the progenitor binary with the back hole at a much closer distance. Assuming this distance to be tidal disruption radius $r_{bt}$, for a massive black hole of mass $M_{bh}=10^6$ $M_\odot$ 
and a binary with semi-major axis $a\sim 1$ $R_\odot$ and total mass $m_\ast \sim 1$ $M_\odot$ we get $r_{bt} = a (M_{bh}/m_\ast)^{1/3}\ll 3$ pc.

\subsection{Survival function}
\label{sec:survival}
If there is no preferred time of ejection, the flight time $t_{f}$ of a HVS of mass $m$ is sampled according to 
\begin{equation}
t_{f} = t_{L}(m)(1-\epsilon_1)\epsilon_2,
\end{equation}
where $\epsilon_1, \epsilon_2$ are random variables uniformly distributed between $0$ and $1$, and $t_{L}(m)$ is the stellar lifetime \citep{Marchetti2018}. In our implementation this is taken to be equal to the main sequence lifetime, modelled according to \cite{Hurley2000a}. 

The probability density function of the variable $t_{f}$ is found to be equal to

\begin{align}
f (t_{f}|m) = -\frac{1}{t_{L}(m)}log\left(\frac{t_{f}}{t_{L}(m)}\right).
\end{align}

Note that the average value of $t_f/t_L$ is then expected to be $0.25$, i.e. on average HVS fly for a quarter of their lifetime. The function $g(t_{f}, m)$ is then the corresponding survival function:
\begin{align}
g(t_{f}, m) =  1-\int_0^{t_f} f (t|m)\, dt = 1 - \frac{t_f}{t_L(m)} + \frac{t_f}{t_L(m)} \log \left( \frac{t_f}{t_L(m)}\right),
\end{align}
for $t_f < t_L(m)$.

\subsection{Galactic potential}
\label{sec:pot}
We model the Milky Way gravitational potential as the sum of four components: central black hole,  bulge,  disc and dark matter halo. Depending on the symmetry, we use Cartesian $(x, y, z)$, spherical $(r, \theta, \phi)$ or cylindrical coordinates $(R, \varphi, z)$. In all three cases we position the GC at the origin and the $z$ axis perpendicular to the Galactic disc. 

The first component is a simple Keplerian potential and it is meant to describe the massive black hole at the centre of the Galaxy with a mass of $M_{bh}=4\times10^6$ M$_\odot$ \citep{Eisenhauer2005, Ghez2008},

\begin{align}
\Phi_{\text{BH}} (r, \theta, \phi) = -\frac{G M_{bh}}{r}.
\end{align}

The second and third components are an Hernquist spheroid \citep{Hernquist1990} and a Miyamoto-Nagai disc \citep{Miyamoto1975} respectively. This form and model parameters are chosen because they are commonly used to parametrize the baryonic components of the Galactic potential in similar studies \citep[e.g.,][]{Johnston1995, Price-Whelan2014, Rossi2016}.

\begin{align}
\Phi_{\text{Bulge}}(r, \theta, \phi) = - \frac{G M_b}{a_b + r},
\\
\Phi_{\text{Disc}} (R, \varphi, z) = -\frac{G M_{d}}{\sqrt{R^2 + \left( a_d + \sqrt{z^2+b_d^2}\right)^2}}.
\end{align}

We use the values $a_b=0.7$ kpc, $M_b = 3.4 \times 10^{10}$ $\text{M}_{\sun}$, $a_d = 6.5$ kpc, $b_d = 260$ pc, $M_d=10^{11}$ $\text{M}_{\sun}$ from \cite{Price-Whelan2014}. 

The last component of the potential is a spheroidal NFW density profile \citep{Navarro1997} and it models the dark matter halo of the Milky Way,

\begin{align}
\rho_{\text{NFW}}(x, y, z) = \frac{M_s}{4\pi r_h^3}\frac{1}{(\xi/r_s)(1+\xi/r_s)^2}, && \xi^2 = x^2 + y^2 + \frac{z^2}{q^2}.
\end{align}

Notice that for the sake of simplicity, in our parametrization $q$ corresponds to the dimensionless axis ratio of our spheroid. The potential associated to this matter density is found by solving Poisson's equation,

\begin{align}
\nabla^2 \Phi_{\text{NFW}} = 4 \pi G \rho_{\text{NFW}}.
\end{align} 

In this study we will focus on three different fiducial Galactic haloes (see Table~\ref{tab:pot}). For model A the chosen values for $M_s, r_s$ are the best fit parameters to the rotation curve of the Milky Way for a spherical halo \citep{Rossi2016}. In model B  we consider an oblate spheroid as variation of this model, and in model C we consider a spherical halo with a significantly different scale radius and mass.

\begin{table}
	\caption{Choice of the NFW scale parameters $M_s, r_s$ and axis-ratio $q$ for the three fiducial haloes used in this work to model the dark matter distribution of the Galaxy.}
	\label{tab:pot}
	\begin{tabular}{l|c|c|c}
		Model & $M_s$ & $r_s$ & $q$ \\
		\hline
		A & $0.76 \times 10^{12} \text{ M}_\odot$ & $24.8$ kpc & 1 \\
		B & $0.76 \times 10^{12} \text{ M}_\odot$ & $24.8$ kpc & 3/2 \\
		C & $1 \times 10^{12} \text{ M}_\odot$ & $20.0$ kpc & 1 \\		
	\end{tabular}
\end{table}

\subsection{Mock catalogues}
\label{sec:mock}
We follow a procedure similar to the one detailed in \cite{Marchetti2018} to generate mock HVS Galactic populations ejected through the Hills mechanism and simulate the effect of the \Gaia selection function. In the same paper, we estimated the current Galactic population of HVSs produced through the Hills mechanism to include $10^5$ members. An ejection sample of this size is therefore generated by sampling the distribution $\mathcal{R}(\bw)$ in  eq.~\ref{eq:R} using a Markov chain Monte Carlo method implemented through the python library \textsc{emcee} (\cite{Foreman-Mackey2013}, based on \cite{Goodman2010}). The sample is then propagated numerically by the software \textsc{galpy} \citep{Bovy2015a}, through the three fiducial models of the Galactic potential presented in Sec.~\ref{sec:pot}, using a time step $\delta t = 0.01$ Myr. The integration time, i.e. the flight time, for each star is dictated by the formulas presented in section~\ref{sec:survival}. 
At the end, the photometric properties of the stars are simulated using the stellar models provided by \cite{Hurley2000a}, the BaSeL SED Library 3.1 \citep{Westera2002} and a map of the dust reddening in the Milky Way presented in \cite{Bovy2016a} \citep[a combination of ][]{Drimmel2003, Marshall2006, Green2015a}. The magnitude in the \Gaia band GRVS is then computed using the polynomial fitting functions provided by \cite{Jordi2010}.

From this catalogue we define a golden sample by imposing two conditions. First, only stars brighter than the 16th magnitude in the $G_{RVS}$ band are selected. This cut filters objects for which \Gaia is expected to measure the line of sight velocity \citep{Cacciari2016, Katz2018}. The second condition that we impose is related to the velocity of the objects appearing in the sample: we impose a total velocity at present time in the Galactic reference frame higher than $450$ km/s. This threshold filters objects which will be clearly recognizable as high velocity  -- i.e. faster than three times the one dimensional Galactic velocity dispersion \citep{Battaglia2005, Brown2010, III2015}. At the end of this selection process our golden samples contain $195, 192, 211$ objects for haloes A, B, C respectively.\footnote{Since we have modelled our analytical $\mathcal{R}(\bw)$ after the MC method we used in \cite{Marchetti2018}, it is not surprising that the size of our golden samples agrees with the one found in the aforementioned paper.} 

We stress that this simulation population is different in nature from the observations reported in the literature \citep[e.g., ][]{Brown2007}. We restrict ourselves to main sequence stars located at Galactocentric distances $\lesssim 20$ kpc due to the limited \Gaia~horizon. For additional information about the catalogue and its construction we refer the reader to \cite{Marchetti2018}. 

\section{Distribution function}
\label{sec:methods}
We study the distribution of HVSs in the configuration space labelled by $\bw = (\mathbfit{x}, \mathbfit{v}, m)$, where $(\mathbfit{x}, \mathbfit{v})$ is the usual phase space coordinate (position, velocity) and $m$ is the stellar mass. We then introduce the density function of HVSs in this space at a time $t$:

\begin{align}
f(\bw; t) = \frac{dN(t)}{ d^3 v \, d^3 x \, dm}.
\end{align}

In this expression, $dN(t)$ represents the number of HVSs in the volume $d^3 v \, d^3 x \, dm$. We now aim to write down this distribution as a combination of two other functions:  the ejection rate distribution $\mathcal{R}(\bw)$, which parametrizes the density of HVSs ejected at a given position of the configuration space per unit time, and the survival function $g(t, m)$, which quantifies the fraction of stars of mass $m$ which survives for at least a time $t$ after ejection. In Sec.~ \ref{sec:mock} we have provided two examples of how these functions might be defined. Notice that we assume a stationary process for the creation of HVSs, meaning that $\mathcal{R}(\bw)$ is not a function of time. 

These definitions allow us to write down the total number of HVSs present in the Galaxy at a time $t$ after the formation of the Milky Way or, equivalently, when the first HVS was ejected:

 \begin{align}
 N(t) = \int d^7 w \,\int^{t}_{0} dt' \,\mathcal{R}(\bw)\, g(t-t', m),
 \label{eq:N}
 \end{align}

In this expression we integrate $\mathcal{R}(\bw)$ over the entire configuration space and over every possible ejection time $t'$. In the last integral, the weight function $g(t-t', m)$ accounts for the fact that not all stars ejected at a time $t'$ will still be alive after a time $t-t'$. 

From the expression for $N(t)$ we can derive the density function by applying a Dirac delta function in configuration space:

\begin{align}
	f(\bw; t) &= \int d^7 w' \int^{t}_{0} dt' \, \mathcal{R}(\bw') \,\times g(t-t', m) \,\delta\left(\mathbfit{W}(\bw', t'; t)-\bw\right).
	\label{eq:finitial}
\end{align}

In this expression we introduced the solution of the equations of motion in configuration space, $\mathbfit{W}(\bw', t'; t)$, which maps the initial condition $\bw'$ at a time $t'$ to the phase space position $\mathbfit{W}$ at a time $t>t'$. The delta function imposes that objects in the position $\bw$ at a time $t$ must have been generated inside the appropriate orbit $\mathbfit{W}(\bw', t'; t)$ at the appropriate time. Note that if we assume that the stellar mass is not a function of time, Liouville's theorem ensures that the map $(\bw', t') \leftrightarrow (\mathbfit{W}, t)$ is bijective and conserves the volume $d^7w$. Because of this, applying the Dirac delta in the integral over $d^7w'$ does not introduce a Jacobian term despite the argument of the Dirac delta not being a trivial function of $\bw'$.  Furthermore, because Hamilton's equations for a single HVS are time invariant, we can write $\mathbfit{W}(w', t'; t) \equiv \mathbfit{W}(w'; t-t')$.  In conclusion, we derive the following:

\begin{align}
f(\bw; t) &= \int^{t}_{0} dt'\, \mathcal{R}(\bw'(\bw; t-t')) \,g(t-t', m).
\label{eq:f}
\end{align}

In this expression we have introduced the trajectory $\bw'(\bw; t-t')$ which is a solution of the argument of the delta function in  eq.~\eqref{eq:finitial} and it can be found by integrating numerically back in time the equations of motion from the starting point $\bw$.

Notice that this final result is completely general: it does not discriminate between bound or unbound objects and can be applied to a variety of ejection mechanisms and lifetime models.

In this analysis we are interested in exploring how the distribution in  eq.~\eqref{eq:f} is affected by the Galactic potential. The dependence on the dynamics is not made clear from the expression itself, but it is hidden in the backwards trajectory $\bw'(\bw; t-t')$ . If we model the Galactic potential using a set of parameters $\btheta$, we can write down the parametric configuration space distribution as:

\begin{equation}
f(\bw; t|\btheta) = \int^{t}_{0} dt'\, \mathcal{R}(\bw'(\bw; t-t'|\btheta)) \,g(t-t', m).
\end{equation}

We can then assign for every value of this parameter vector a likelihood to the observation of $\NHVS$ HVSs in the configuration space points $\{\bw_1, \ldots, \bw_\NHVS\}$ at a time $t$:

\begin{align}
\mathcal{L}(\btheta) = \sum_i^\NHVS f(\bw_i; t|\btheta).
\label{eq:L}
\end{align}

While the likelihood function formally depends on the observations, in order to simplify the notation our expression does not make this dependence of $\mathcal{L}$ on $\bw_i$ explicit.

Our implementation is strictly a forward-fitting algorithm, meaning that it does not produce model-independent results, but it can be used to constrain any parametric model. The first obvious advantage of this technique is that it allows us to parametrize (hence fit) any aspect of the HVS population. For example, we could easily use an observed sample to constrain a parametric version of $\mathcal{R}(\bw)$. In this case, we would write the dependence on model parameters explicitly into its expression. Notice however that, in order to compare different ejection mechanisms, the rates $\Gamma$ should be fixed or at least be left as free parameters. Secondly, we stress that the technique described here can be implemented for unbound and bound trajectories alike. The periodicity is not an issue thanks to the explicit time dependence of $g(t_f, m)$. The presence of this function in the integral also means that the time integration should be performed only between now and a time $t_L(m)$ in the past, since $g(t_f, m)$ is zero by design after this point. Thirdly, because the stars are tested individually and not as a sample, a single one is able to rule out any Galactic potential not consistent with Galactocentric origin or any ejection model unable to reproduce the range of allowed initial velocities.

In practice, the evaluation of $\mathcal{L}(\btheta)$ is performed by integrating numerically back in time the HVS orbits from the observed positions $\bw_i$ under the influence of the potential specified by $\btheta$. For consistency, our set-up matches the one employed for the creation of the mock catalogue. Given the orbit $\bw(t-t')$ as a function of the backwards time coordinate $t'$ we can then evaluate the integral in  eq.~\ref{eq:f} in the configuration space volume where $\mathcal{R}(\bw)$ is non-zero. Because of the presence of the Dirac deltas this volume is formally a $4-$d  space embedded in the $7$-d configuration space. To perform the integral and account for numerical errors we swap the Dirac deltas with Gaussian kernels calibrated against the numerical precision of the orbit back-propagation code and truncated at $4$ standard deviations. This introduces two smoothing parameters which correspond to $\sigma_r=10$ pc and $\sigma_L=10$ km $\times$ pc /s. For a physical justification of these values we refer the reader to Appendix~\ref{sec:stability}.

\section{Likelihood function}
\label{sec:res}

To test our method, we study the likelihood $\mathcal{L}(\btheta)$ for the golden sample of HVSs simulated in Sec.~\ref{sec:numsim} as input data and the halo potential parameters as variable $\btheta$. In Sec.~\ref{sec:L1} we explore the parameter space $\btheta = (M_s, r_s)$, while keeping $q$ fixed at the fiducial value; and in Sec.~\ref{sec:L2} we assume $\btheta = (q)$ and freeze $M_s, r_s$. We then discuss in Sec.~\ref{sec:corr} the implications for the full parameter space $\btheta~=~(M_s, r_s, q)$. We use the subscript $0$ (e.g. $q_0$) to indicate the fiducial value of our halo parameters.

This choice of parameter space allows us to quantify in a general way how precisely HVSs can constrain the mass and shape of the Galactic dark matter halo. Notice in particular that while the geometry of the dark matter halo might be non-trivial \citep[e.g., ][]{Law2010}, exploring the resulting high-dimensional parameter space is outside the scope of this work. Since the shape of a halo is, in general, not expected to be quantifiable with a simple parameter, in Appendix~\ref{sec:triaxial} we discuss the constraints on a triaxial configuration.

Our analysis also helps us identifying which stars are particularly suited to measure the Galactic halo, see Sec.~\ref{sec:prospects} where we characterize the observational properties of this sample.

\subsection{Likelihood in $M_s-r_s$ plane}
	\label{sec:L1}
\begin{figure}
			\includegraphics[width=0.5\textwidth]{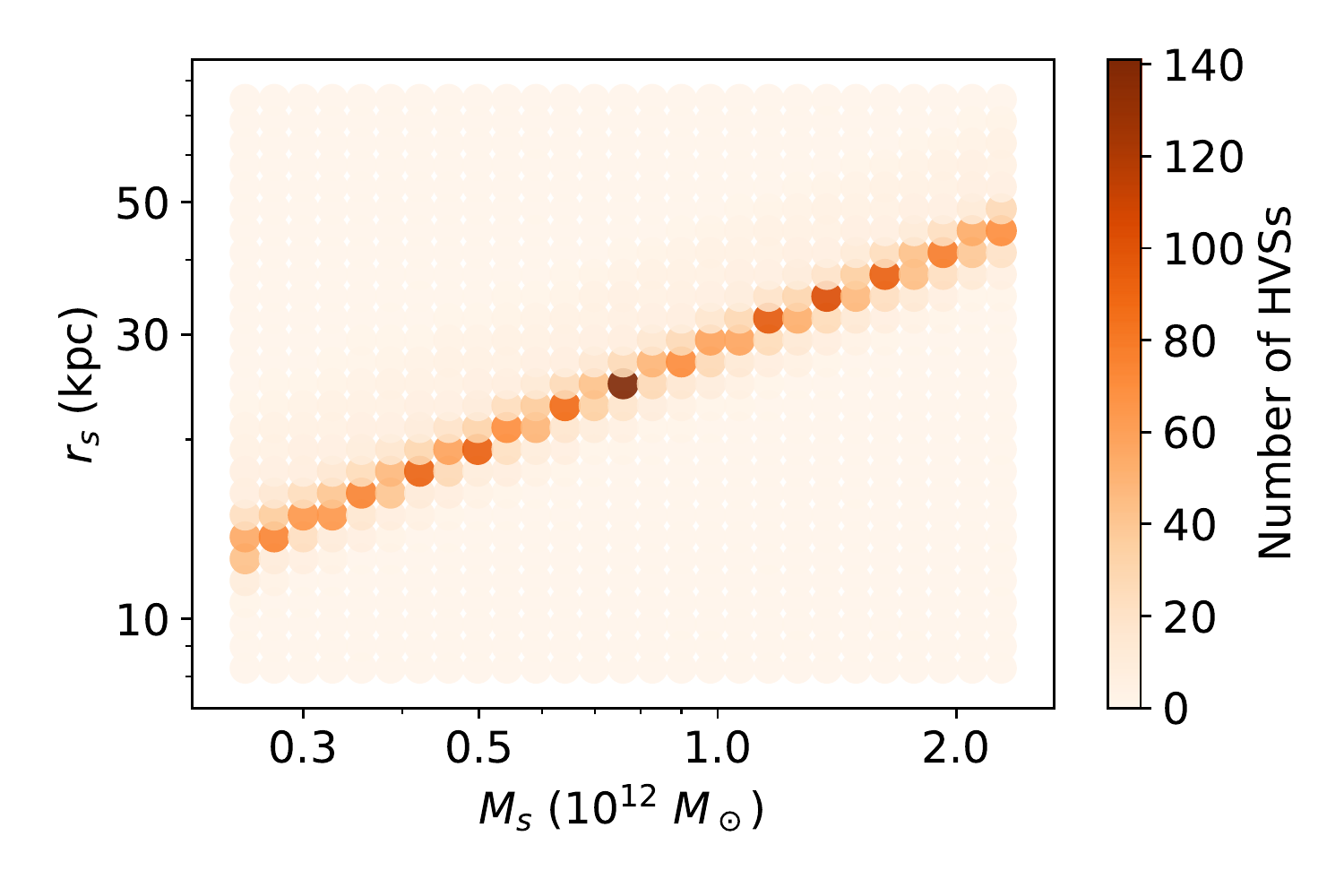}
			\caption{Number HVSs with non-zero likelihood for potentials defined in the plane $M_s-r_s$with a constant shape parameter $q$ set at its fiducial value. Here we consider only the average constrainers (see Sec.~\ref{sec:L1}). The peak corresponds to the fiducial model A, under which these stars were propagated. The clear degeneracy line corresponds to a constant value of $\alpha = M_s/r_s^2$.}
			\label{fig:N1}
		\end{figure}

\begin{figure}
		\includegraphics[width=0.5\textwidth]{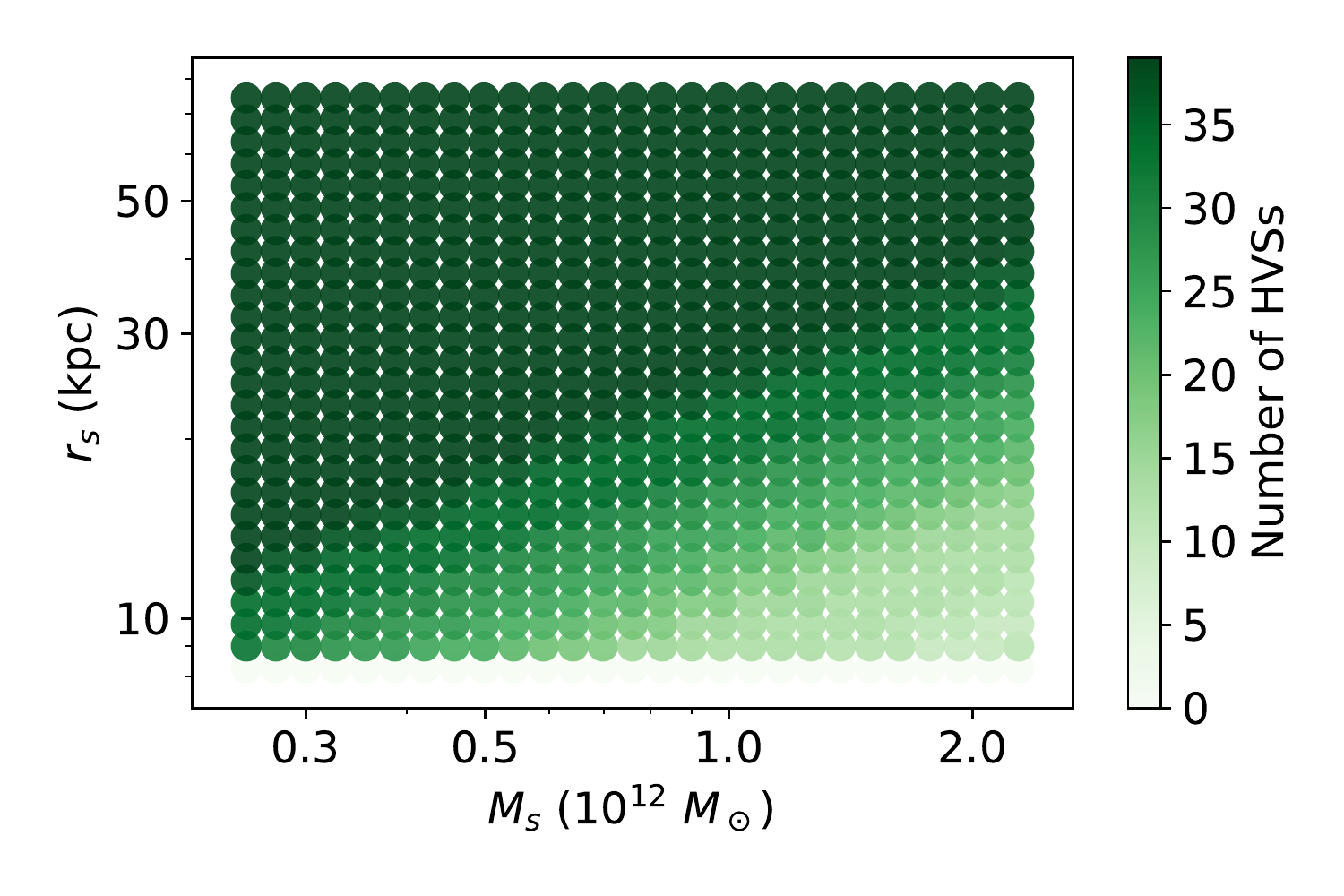}
		\caption{Number of HVSs with non-zero likelihood for every potential explored in the plane $M_s-r_s$. Same as Fig.~\ref{fig:N1}, but here we consider only the poor constrainers (see Sec.~\ref{sec:L1}). No peak is visible for the fiducial model A, under which these stars were propagated.}
		\label{fig:N2}
	\end{figure}

		For the three haloes A, B, C we evaluate the likelihood,  eq. \eqref{eq:L}, in the space $r_s, M_s$ using a coarse grid of size $27\times27$.  Based on the results, we define three classes of HVSs in our golden sample: strong, average and poor constrainers. This classification is based on the number of points on our grid with non-zero likelihood. Fig. \ref{fig:N1} and Fig. \ref{fig:N2} depict the significantly different trend of the latter two classes for halo A. The strong constrainers (not shown) are stars for which no particular trend in the likelihood was identified and have non-zero likelihood only in the fiducial model. 
		
		For every star we call $n$ the number of non-zero likelihood points associated to it: strong constrainers have $n=1$ and average constrainers have $1\leq n\leq300$. The value $300$ is picked from visual inspection of the individual likelihoods. 

		From Fig. \ref{fig:N1} we infer that HVSs are sensitive exclusively to the parameter
		\begin{equation}
			\alpha = \frac{M_s}{r_s^2}
		\end{equation}
		 in the $M_s - r_s$ plane. Following the established notation, we call $\alpha_0$ the fiducial value of this parameter. Notice that, for a spherical NFW potential, this degeneracy is natural and every $\alpha$ corresponds to a value of the local force at small radii:
		 \begin{align}
		 	F \propto \frac{M(<r)}{r^2} = \frac{1}{r^2} \int_0^{r} 4 \pi y^2 \rho_{\text{NFW}}(y) \, dy \approx \frac{M_s}{2r_s^2},
		 \end{align}
		 where we expanded the integral around  $r/r_s=0$. A simple physical interpretation of this degeneracy is that the innermost region of the halo is responsible for the majority of the deceleration experienced by these HVSs.
		 
		 For any single star we can interpolate the likelihood in our coarse grid and obtain an estimate of the $1\sigma$ error on $\alpha$ associated to it, which we call $\sigma_\alpha$.  Fig. \ref{fig:scalingalpha} shows how $n$ and $\sigma_\alpha$ are related to each other. The scaling $\sigma_\alpha \propto \sqrt{n}$ is indicative of the fact that a constant $\alpha$ represents a $1$d curve in the $2$d $M_s-r_s$ plane. After confirming the absence of bias in the measurement of $\alpha$ for the individual stars, we estimate the $1\sigma$ error of the stacked likelihood by assuming normality and using the geometric mean of the individual variances:

		\begin{figure}
			\includegraphics[width=0.5\textwidth]{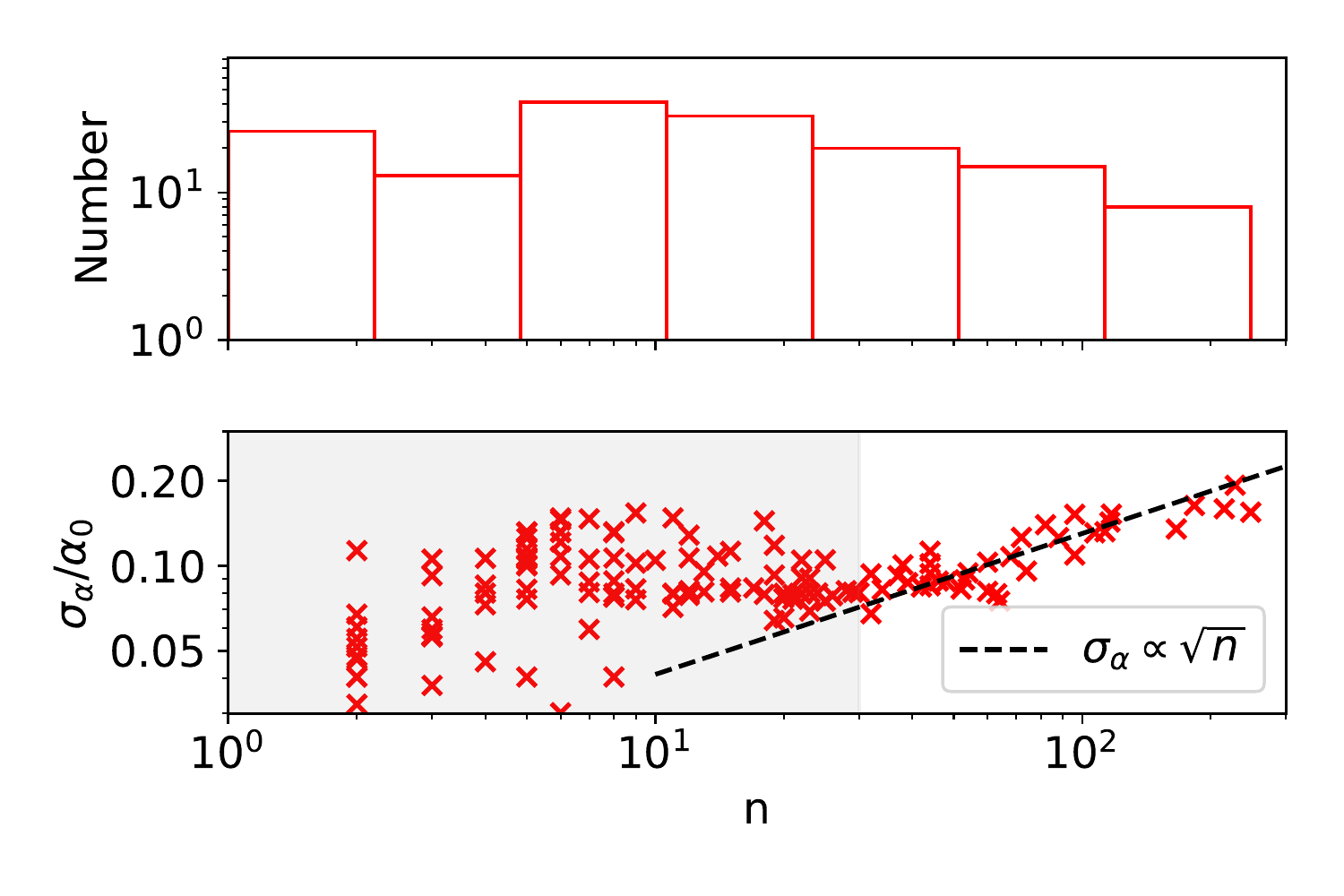}
			\caption{Relation between $n$ (number of potentials with non-zero likelihood in the plane $M_s-r_s$) for stars propagated in halo A and the $1\sigma$ error on $\alpha=M_s/r_s^2$. The top histogram shows the distribution of $n$ for the stars in our golden sample. The relation saturates for  $n\lesssim 30$, when grid effects start to hinder the estimate of $\sigma_\alpha$. Therefore, the relation $\sigma_\alpha(n)$ is calibrated using only points outside the shaded area.}
			\label{fig:scalingalpha}
		\end{figure}

		 \begin{equation}
		 	\hat{\sigma}_{\alpha, q_0} = \left(\sum_i \frac{1}{\sigma^2_\alpha(n_i)}\right)^{-1/2}  \propto  \left(\sum_i \frac{1}{n_i}\right)^{-1/2} 
		 \end{equation}
		 where the proportionality constant for $\sigma_\alpha(n)$ is fitted independently for our three halo models and, for halo A, it is presented in Fig. \ref{fig:scalingalpha}. The variable $n_i$ represents the $n$ corresponding to the $i-$th star. The proportionality constant in $\sigma_\alpha(n)$ is found to be equal to $(1.7, 1.2, 3.3)\times10^{7}$ M$_\odot$ $kpc^{-2}$ for halo A, B, C respectively. 
		 
		 Table~\ref{tab:classes} reports the number of sources in each class for our three fiducial haloes and the effective precision in $\alpha$ expected from the strong and average constrainers using this method. Notice that the value of the combined errobar $\hat{\sigma}_{\alpha, q_0}$ is dominated by low-$n$ stars, meaning that we are extremely susceptible not only to the inferred $\sigma_\alpha(n)$ but also to changes in the distribution of the variable $n$. To mitigate this effect, the $\hat{\sigma}_\alpha/\alpha_0$ mentioned in the table does not use an extrapolated $\sigma_\alpha(n)$ to values $n<30$, but assumes the constant value $\sigma(n=30)$.
		
		In Sec.~\ref{sec:prospects} we discuss how various orbital properties strongly correlate with the likelihood classification and how this information can be used to guide future detections of HVSs.  
	
\subsection{Likelihood in $q$}
	\label{sec:L2}	
\begin{figure}
		\includegraphics[width=0.5\textwidth]{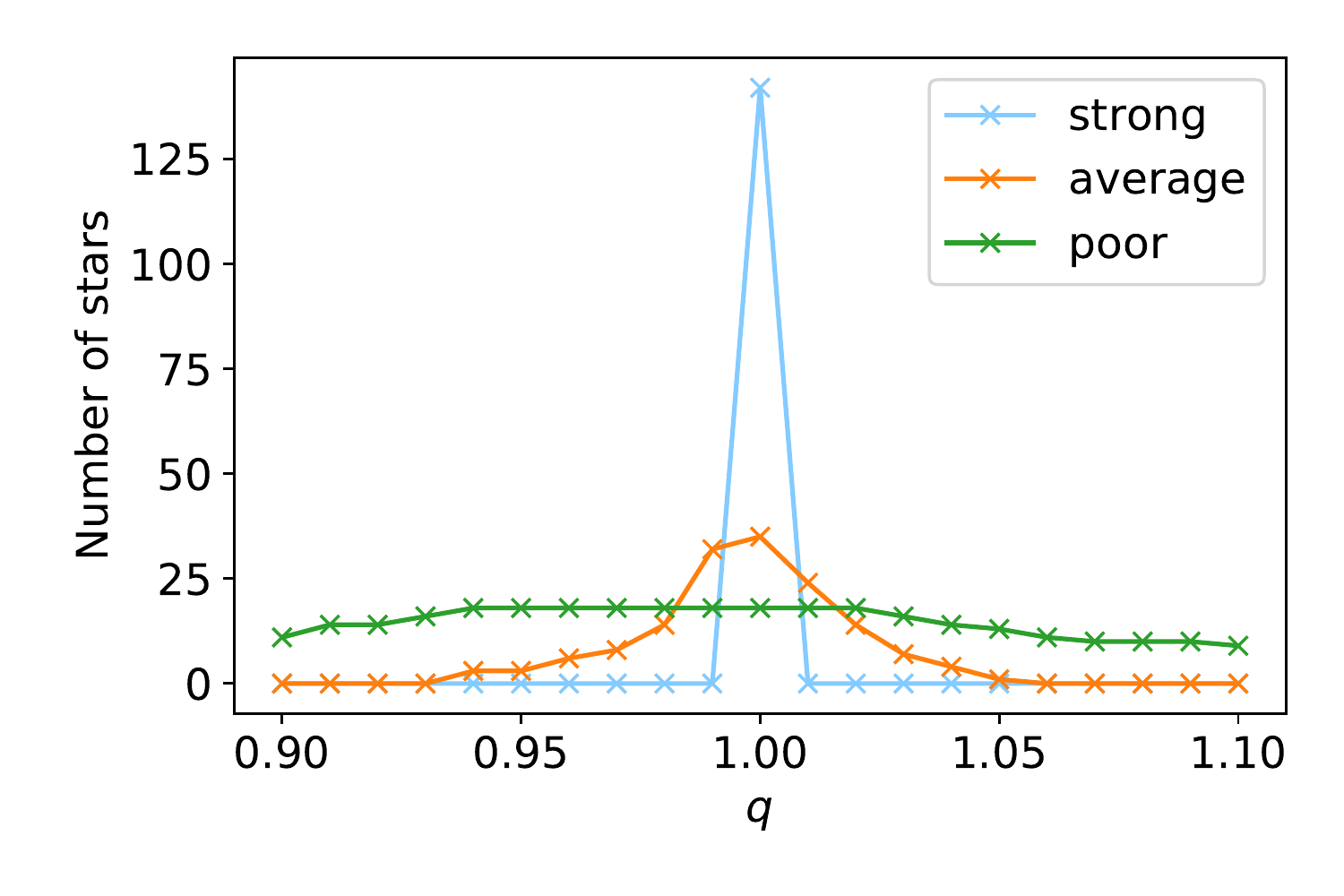}
		\caption{Number of stars with non-zero likelihood under a potential with varying $q$ and fixed $M_s, r_s$. We consider for this figure average, poor and strong constrainers for the shape parameter $q$ (see Sec.~\ref{sec:L2}). The peak corresponds to the fiducial model A, under which these stars were propagated.}
		\label{fig:classesc}
	\end{figure}

	Similarly to what we did for the parameter $\alpha$, we evaluate the likelihood in  eq. \eqref{eq:L} by varying the parameter $q$, while fixing the values of $M_s$ and $r_s$ to their fiducial values (Table~\ref{tab:pot}). We develop again a classification based on the number of non-zero likelihood points, which is shown in Fig.~\ref{fig:classesc}. Notice that while the poor constrainers might prefer the fiducial model, we confirm that their individual likelihoods are either extremely broad or significantly biased -- sometimes excluding the fiducial value at the $3\sigma$ level. 
	
	We stress that the labels we attach to the HVSs (either poor, average or strong constrainer) are independent statements for the two parameters $\alpha$ and $q$. We find, however, significant overlap between them: for halo A, among the $142$ strong constrainers for $q$, $126$ are in the average category for $\alpha$ and $15$ are in the strong one. In fact, the performance of every star for $q$ is always equal or better than for $\alpha$. This implies that HVSs are more sensitive to one parameter than the other in our scheme. This is expected; while the parameters $M_s$ and $r_s$ set the deceleration, a incorrect parameter $q$ can disrupt the ejection point of quasi-radial orbits by introducing additional torque. 
	
	\begin{figure}
		\includegraphics[width=0.5\textwidth]{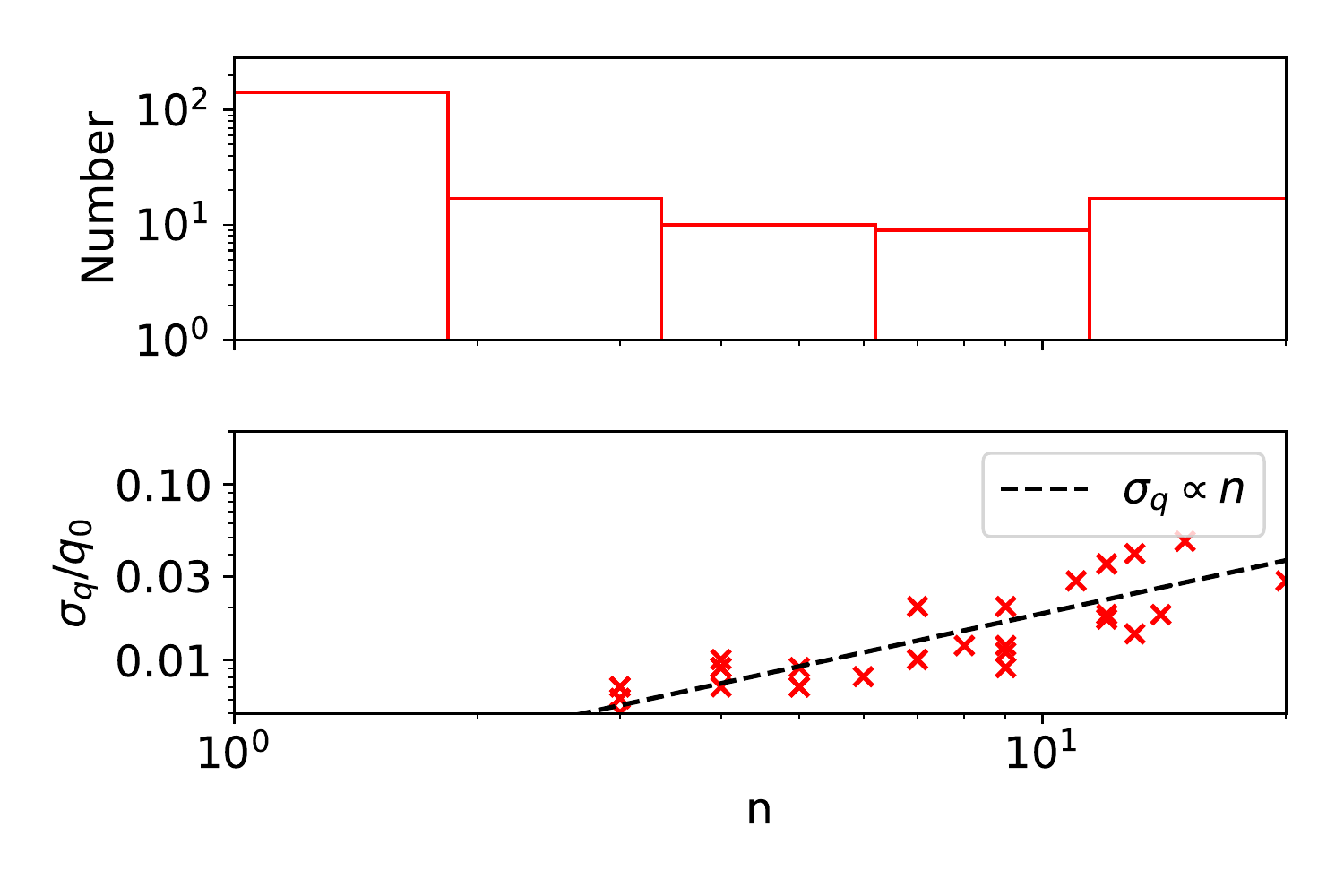}
		\caption{Relation between $n$ (number of potentials with non-zero likelihood for the parameter $q$) for stars propagated in halo A and the $1\sigma$ error on $q$. The top histogram shows the distribution of $n$ for the stars in our golden sample. The relation $\sigma_q(n)$ is calibrated using only a fraction of the sample, presented by the plotted points.}	
		\label{fig:scalingc}
	\end{figure}
	
	For each star, we can relate the number $n$ of non-zero likelihood points for the parameter $q$ to the expected confidence interval $\sigma_q$. Fig. \ref{fig:scalingc} shows how the two are related and provides the distribution of the values of $n$ for the strong and average constrainers for our halo A model (the same trend is observed in all models). As discussed in Sec.~\ref{sec:L1} for the parameter $\alpha$, this scatter plot also fixes the proportionality constant for the stacked uncertainty:

		\begin{equation}
			\hat{\sigma}_{q, \alpha_0} = \left(\sum_i \frac{1}{\sigma^2_q(n_i)}\right)^{-1/2} \propto \left(\sum_i\frac{1}{n_i^2}\right)^{-1/2}.
		\end{equation}

	Notice that this time $\sigma_q(n) \propto n$, since our mesh for this section is constructed on the space of the parameter $q$ directly. As before, because we are extremely susceptible to our reconstruction of $\sigma_q(n)$, we do not extrapolate $\sigma_q(n)$ below the value $n<3$, but we assume a constant value. Notice how bias notwithstanding, some of the poor constrainers can still be used to calibrate $\sigma_q(n)$. The proportionality constant in $\sigma_q(n)$ is found to be equal to $(1.8, 2.3, 2.3)\times10^{-3}$ for halo A, B, C respectively. 
	
	Our results are summarized in Table~\ref{tab:classes}, where we present the estimated precision $\hat{\sigma}_q$ for the combination of our average and strong constrainers.

\begin{figure}
	\includegraphics[width=0.5\textwidth]{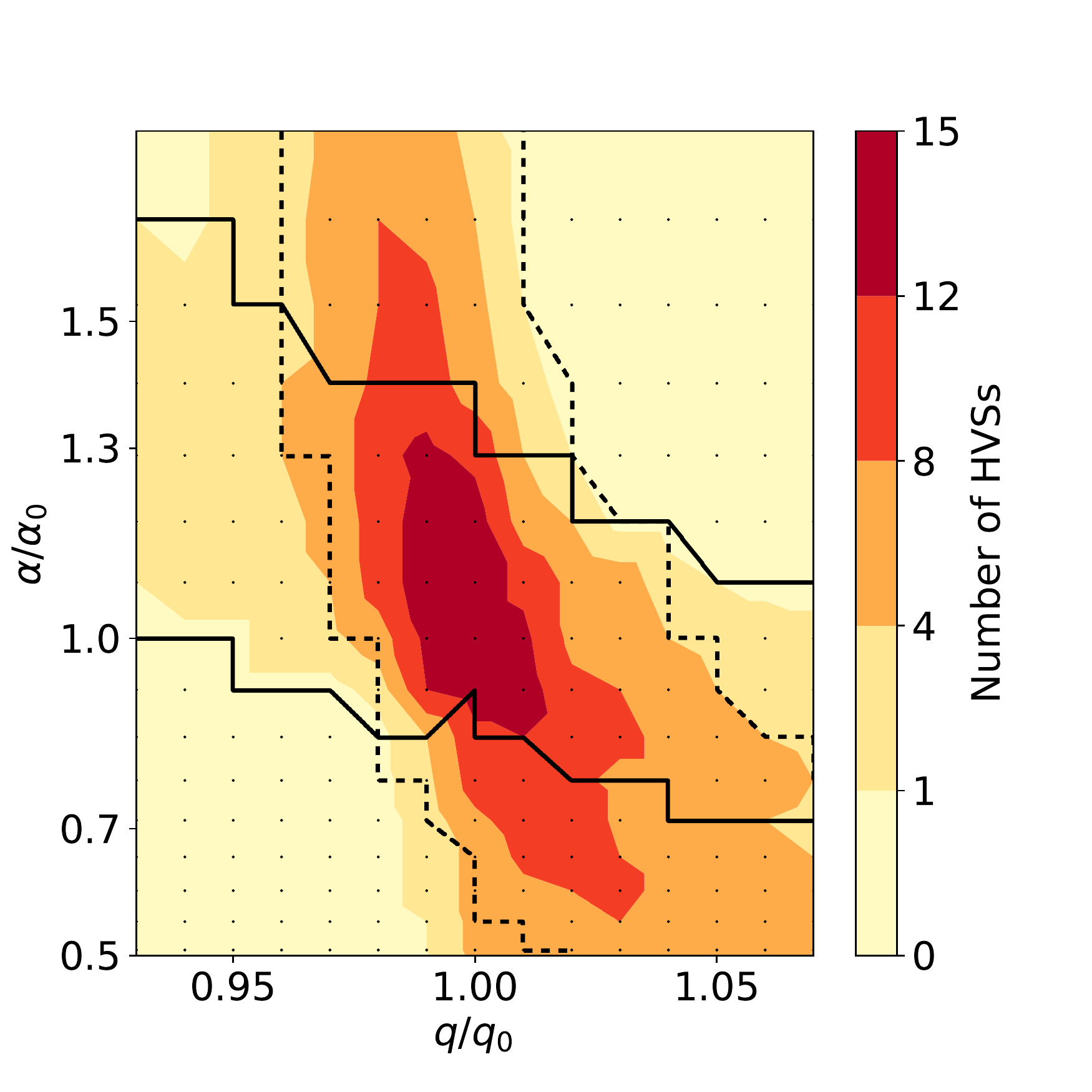}
	\caption{Number of HVSs propagated in halo A for which a given halo, parametrized by the effective scale parameter $\alpha=M_s/r_s^2$ and shape parameter $q$, is allowed.  The peak at $(1, 1)$ marks the fiducial values for halo A. The figure was created using stars for which a visible spread in the likelihood is present in our grid and the result proves that there is correlation between the two parameters. The contours are created by linearly interpolating the values found on a grid. To illustrate the origin of this degeneracy, the dashed and solid lines delimit regions allowed by two particular stars.}
	\label{fig:stack}
\end{figure}

\begin{figure*}
	\includegraphics[width=0.45\textwidth]{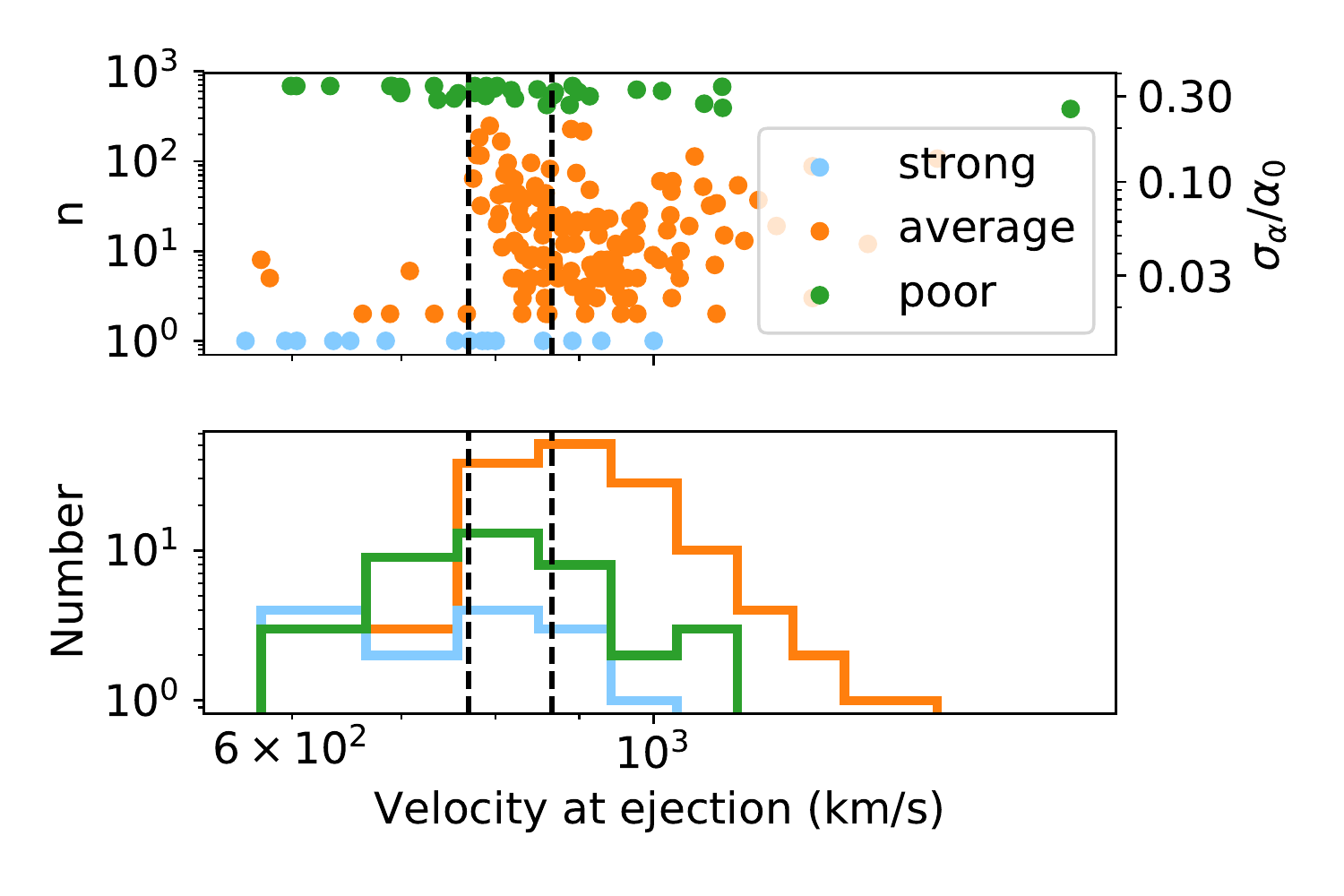}
	\includegraphics[width=0.45\textwidth]{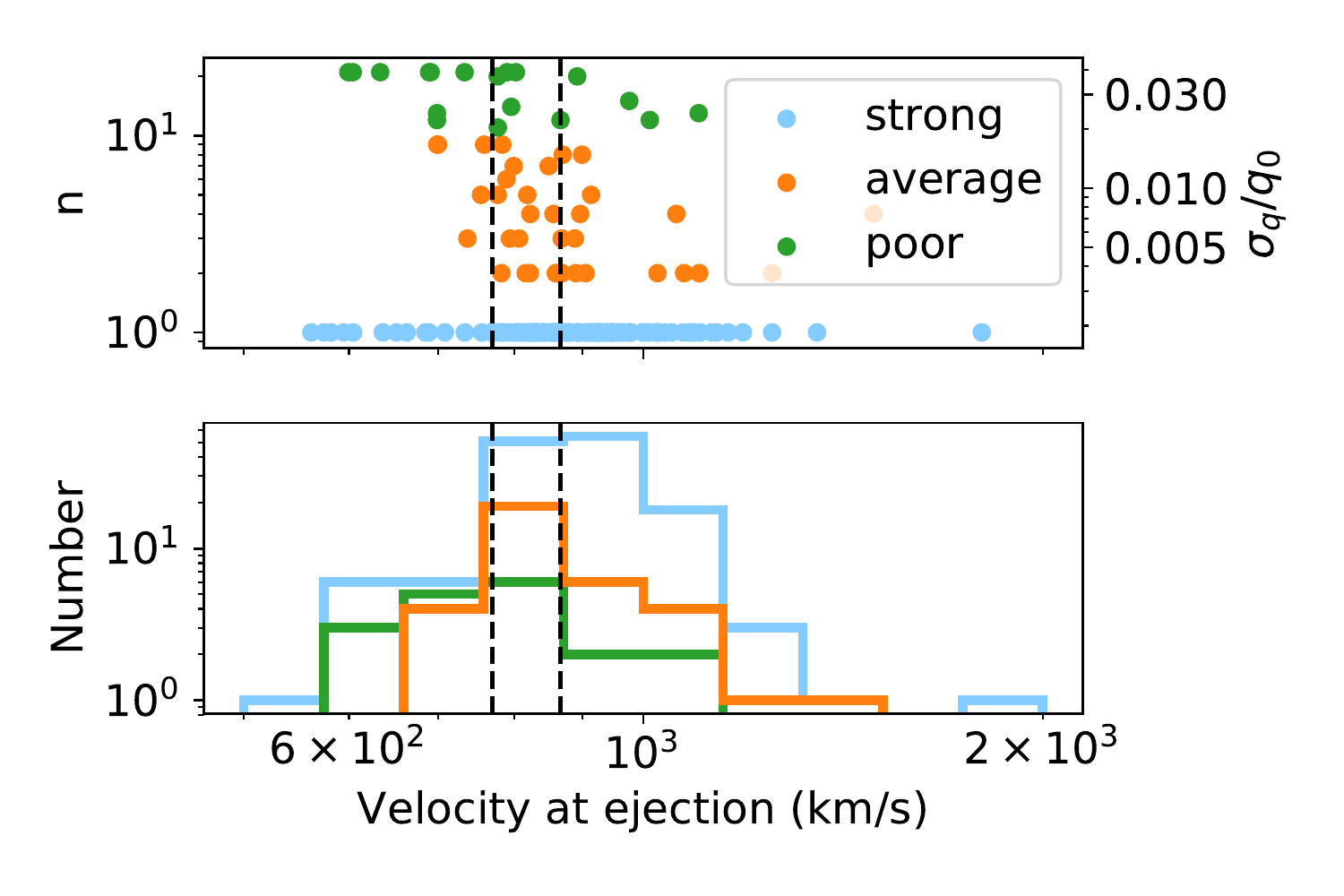}
	
	\includegraphics[width=0.45\textwidth]{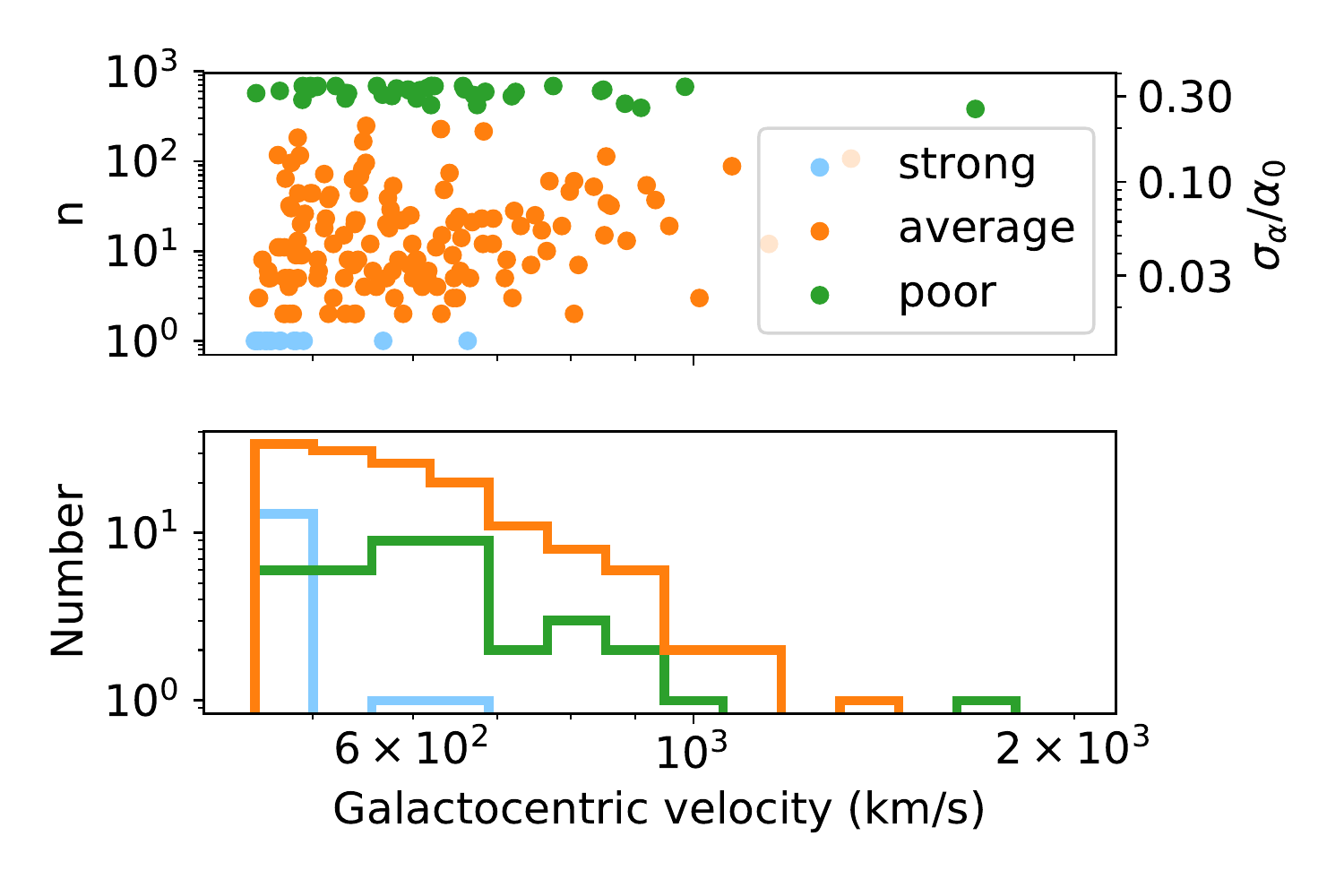}
	\includegraphics[width=0.45\textwidth]{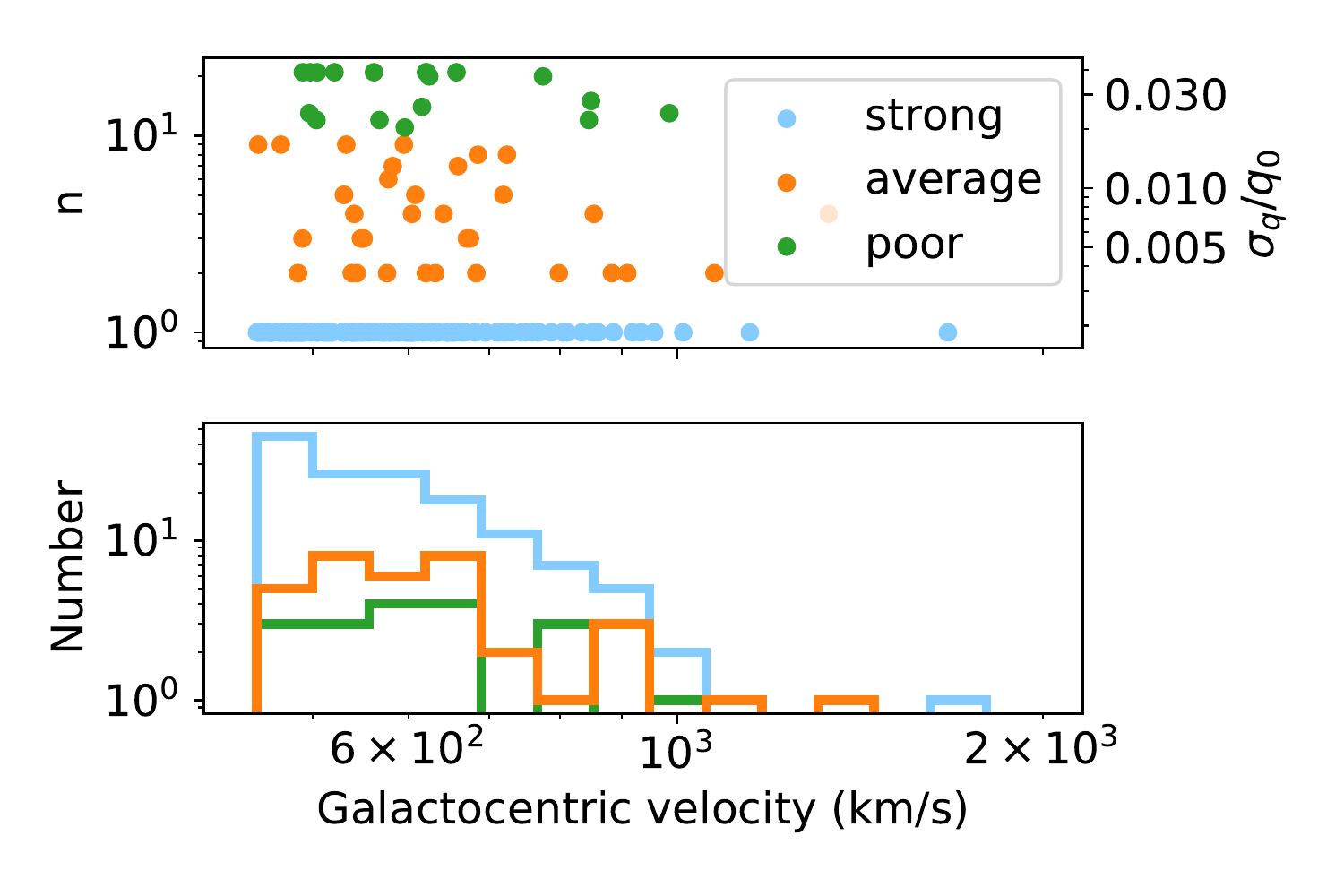}
	
	\includegraphics[width=0.45\textwidth]{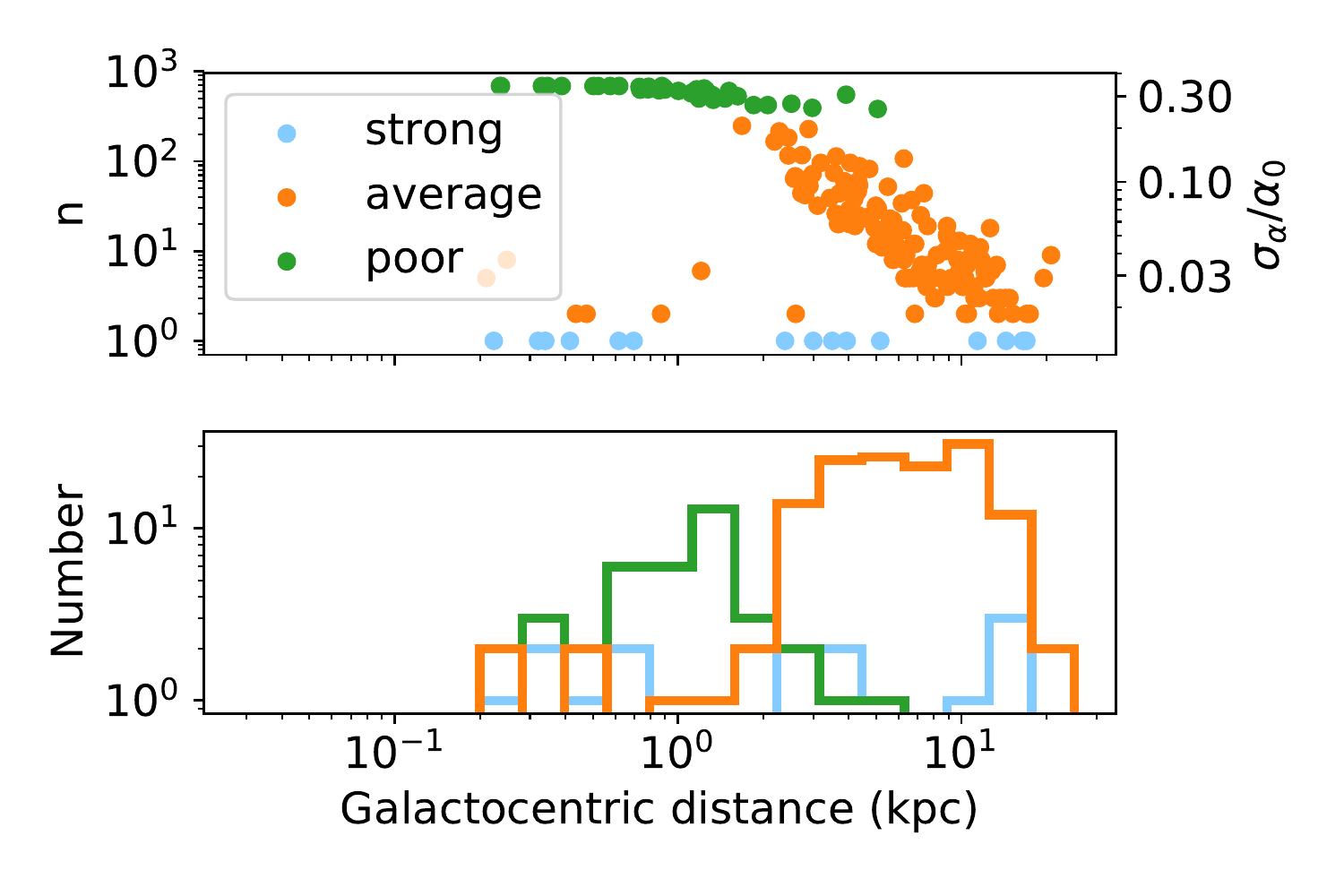}
	\includegraphics[width=0.45\textwidth]{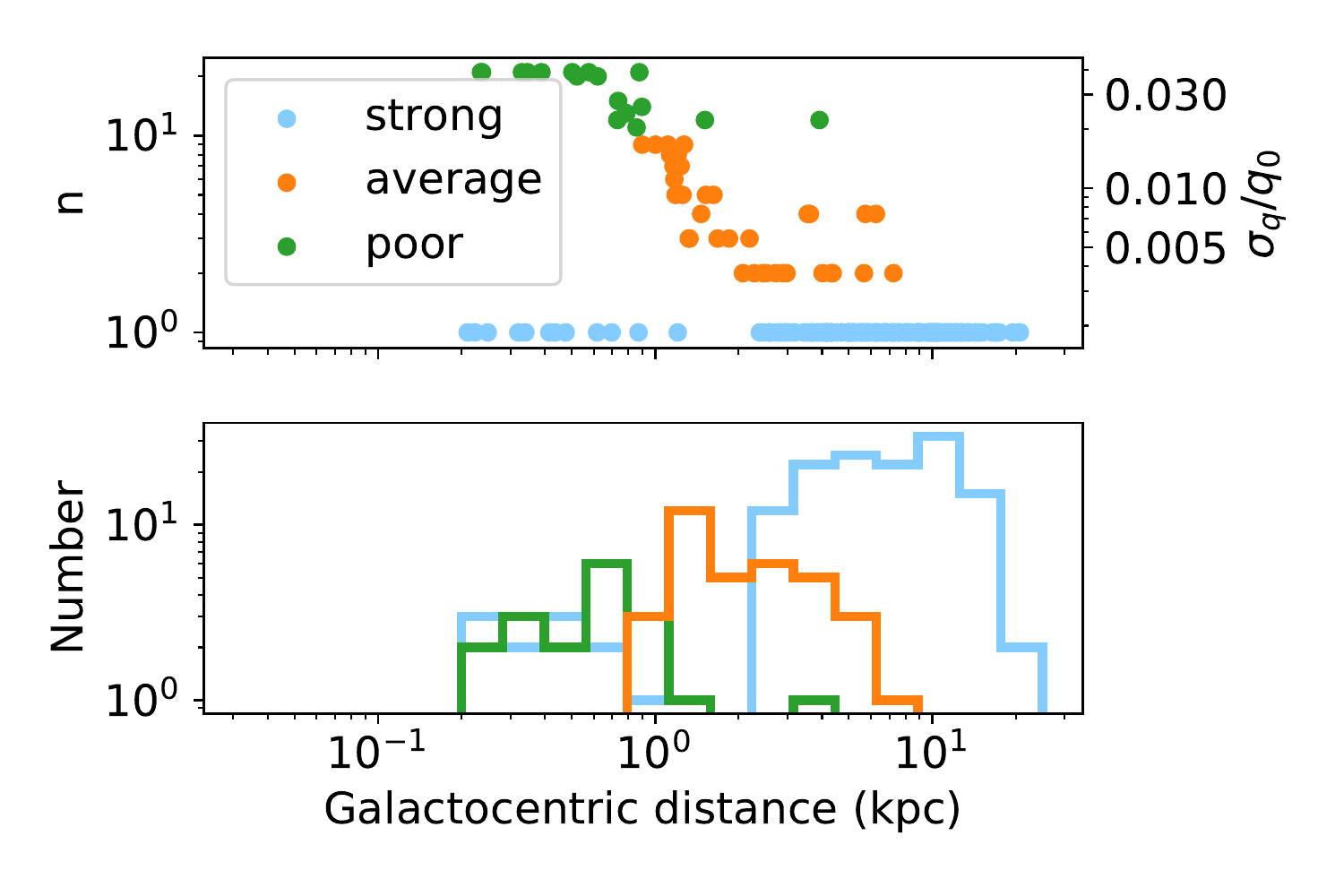}
	
	\includegraphics[width=0.45\textwidth]{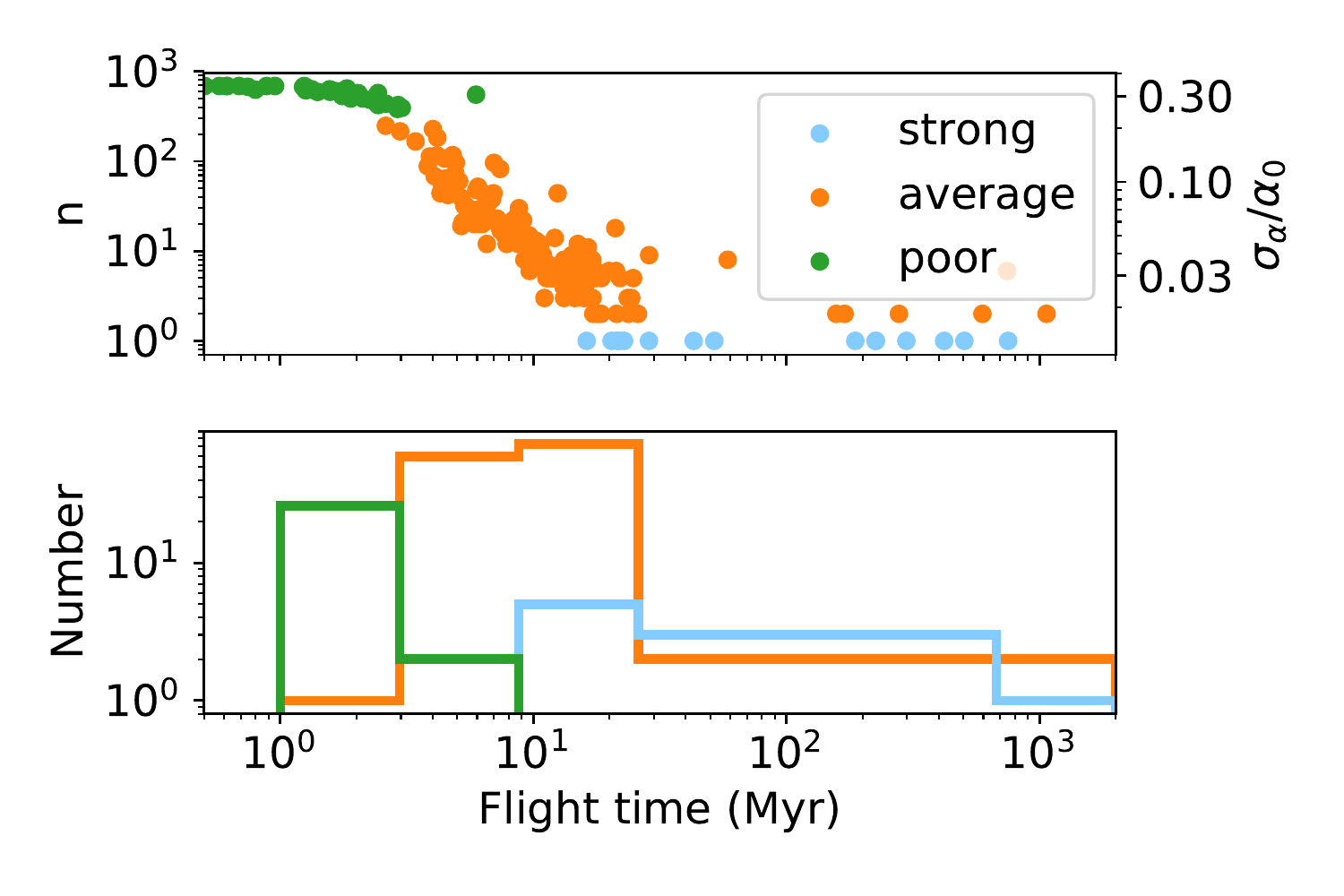}
	\includegraphics[width=0.45\textwidth]{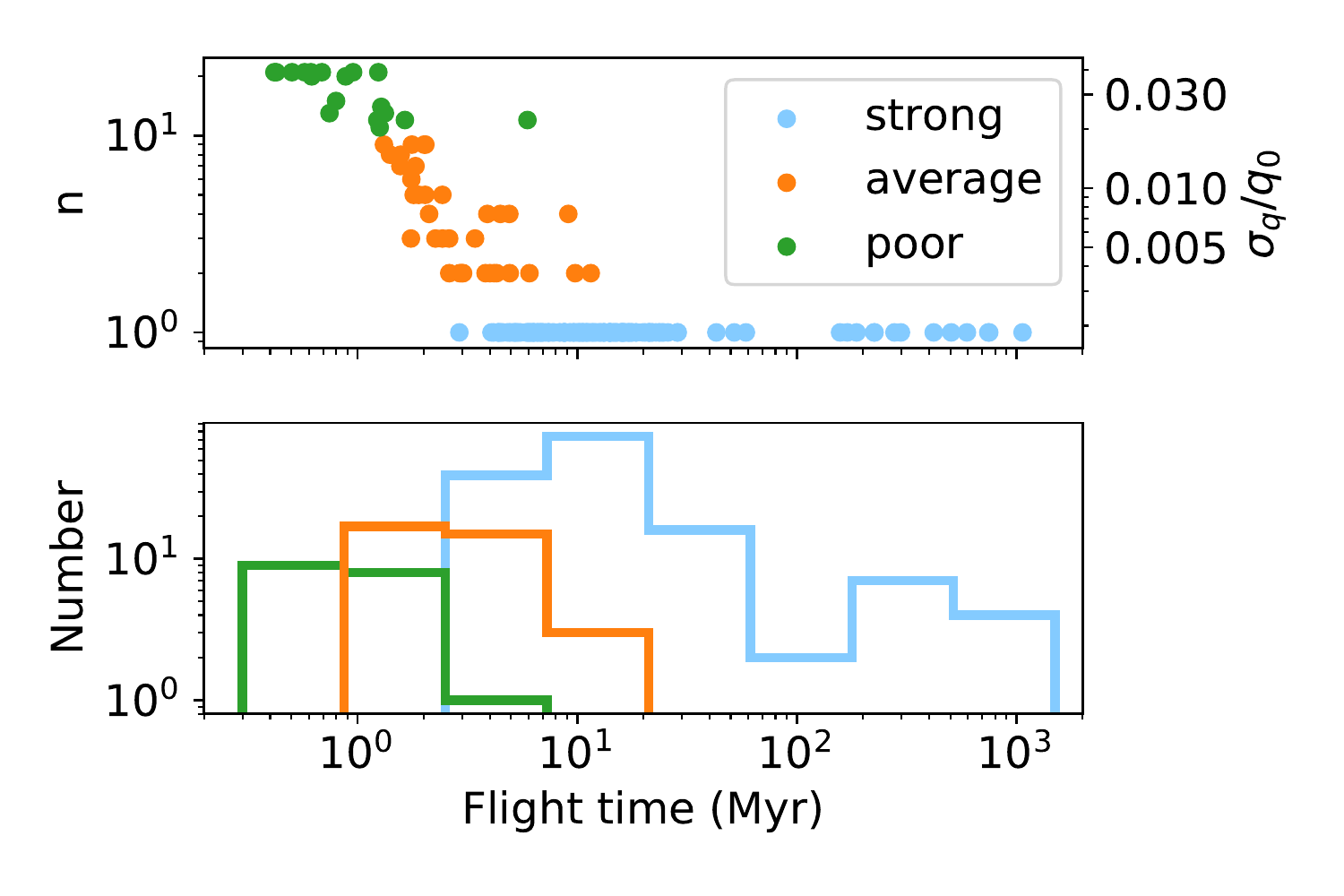}
	\caption{The number of HVSs (Number, histograms in lower panels) and number of non-zero likelihood points per star (n, upper panels) as a function of various kinematic properties. Markers and lines in green, orange and cyan correspond respectively to poor, average and strong constrainers as defined in Sec.~\ref{sec:L1} and \ref{sec:L2}. The results shown are for model A for the Milky Way's dark matter halo, but identical trends are found in model B and C. Overall, these plots show the presence or absence of correlation between the kinematic properties of HVSs and their ability to constrain the parameter $\alpha$ (left column) or $q$ (right column), parametrized by the expected individual relative errors $\sigma_\alpha/\alpha_0$ and $\sigma_q/q_0$. In the ejection velocity plots, the two vertical dashed lines mark, from left to right, the minimum velocity necessary to reach a Galactocentric distance equal to the scale radius $r_s$ and $250$ kpc respectively. }
	\label{fig:correlationsc}
\end{figure*}

\begin{table*}
	\caption{Number of HVSs with different constraining power for the fiducial haloes considered in this work and predictions for the combined relative errors $\hat{\sigma}_\alpha/\alpha_0$ and $\hat{\sigma}_c/\alpha_q$ on the NFW effective parameters $\alpha=M_s/r_s^2, q$. The lower bound on the error $\hat{\sigma}_q$ (or $\hat{\sigma}_\alpha$) is found in Sec.~\ref{sec:L2} (\ref{sec:L1}) by fixing $M_s, r_s$ ($q$) to the fiducial values and exploring only the direction $q$  (plane $M_s-r_s$). The upper bound is found in Sec.~\ref{sec:corr} after estimating the correlation coefficient between the two parameters.}
	\label{tab:classes}
	\begin{tabular}{l|c|c|c|c||c|c|c|c}
		 Model & \multicolumn{4}{c}{$\alpha$} & \multicolumn{4}{c}{$q$} \\
		& \# poor & \# average & \# strong & $\hat{\sigma}_\alpha/\alpha_0$ & \# poor & \# average & \# strong & $\hat{\sigma}_q/q_0$ \\
		\hline
		A & $39$ & $141$ & $15$ &  $0.63\%$ - $0.95\%$ & $18$ & $35$ & $142$ & $< 0.1\%$\\
		B & $6$ & $130$ & $56$ &  $0.47\%$  - $0.71\%$ & $23$ & $38$ & $131$ & $< 0.11\%$\\
		C & $33$ & $143$ & $35$ & $0.64\%$ - $0.96\%$&  $9$ & $32$ & $170$ & $<0.1\%$ \\	
	\end{tabular}
\end{table*}

\subsection{Correlation between $\alpha$ and $q$}
\label{sec:corr}

In Fig. \ref{fig:stack} we show how many stars allow a certain halo model parametrized by $\alpha$ and $q$. To generate this figure, we have explored the whole parameter space $\btheta = (M_s, r_s, q)$ only for $15$ stars in the average category for both $\alpha$ and $q$, as defined in the previous two subsections. We consider only this subset because these stars have a broad likelihood in both projections and are particularly suited to show the presence of correlation. 

A correlation is clearly visible for every star, but while in the plane $M_s-r_s$ they all constrain the same combination $\alpha$, the same is not true in the plane $q-\alpha$. As an example of this, in Fig.~\ref{fig:stack} we also show the degeneracy stripe for 2 stars. Because of this, we expect both direction and size of the combined constraints to depend on the particular selection bias of our sample. 

To give an estimate of the impact of this correlation on our reconstructed errors we assume the combined likelihood to be a bivariate normal distribution. Notice that in the previous two sections we verified that the two one-directional log-likelihoods $\log \mathcal{L}(\alpha, q_0)$ and $\log \mathcal{L}(\alpha_0, q)$ are both normal. However, the $1\sigma$ error bars $\hat{\sigma}_{\alpha, q_0}$ and $\hat{\sigma}_{q, \alpha_0}$ we found in Sec.~\ref{sec:L1},~\ref{sec:L2} do not correspond to the standard deviations of the full $\log \mathcal{L}(\alpha, q)$ in the presence of correlation. A bivariate log-likelihood up to constant terms can be written as:
\begin{align}
&\log\mathcal{L}(\alpha, q) =\\& -\frac{1}{2(1-\rho^2)} \left[ \frac{(c - c_0)^2}{\hat{\sigma}_c^2} + 
 \frac{(\alpha - \alpha_0)^2}{\hat{\sigma}_\alpha^2} - \frac{2\rho(\alpha-\alpha_0)(c-c_0)}{\hat{\sigma}_\alpha \hat{\sigma}_c} \right].
\end{align}

Where $\hat{\sigma}_\alpha, \hat{\sigma}_q$ are the standard deviation for the two parameters and $-1<\rho<1$ is the correlation coefficient. From this expression it is clear that the standard deviations found in Sec.~\ref{sec:L1},~\ref{sec:L2} are an underestimate of the real error bars in the full $\alpha, q$ parameter space and should be multiplied by a factor $(1-\rho^2)^{-1/2}\geq1$.  An estimate of the correlation coefficient $\rho$ can be found by fitting the function in Fig.~\ref{fig:stack} by assuming that the number of stars with non-zero likelihood trace the underlying likelihood contours. By doing this, we obtain $\rho = -0.74$, which corresponds to a factor $1.5$ for the uncertainties. This multiplication provides us with upper limits for the $1\sigma$ errors, as reported in Table~\ref{tab:classes}. Notice that we consider this to be an overestimate of the real uncertainties because the individual contours are in reality non-normal, non-linear and have slightly orthogonal constraints among each others. 

The quoted precisions for $\alpha, q$ in our summary table are remarkable. This is a by-product of the extremely stringent condition that all HVS orbits should be radial and cross the ejection region near the GC, which represents a limited volume of the Galactic phase-space. In our numerical implementation, this volume is determined by the hyper-parameters $\sigma_r$ and $\sigma_L$, which set the maximum distance from the GC, $r$, and the maximum angular momentum, $L$, allowed inside the ejection region. In our testing, relaxing the condition on the angular momentum worsens the constraints in $\alpha, q$ considerably, meaning that the zero-angular momentum condition is the dominant factor that allows HVSs to constrain the NFW profile.

\subsection{Observational prospects}
\label{sec:prospects}

Fig.~ \ref{fig:correlationsc} show the orbital characteristics of the strong, poor and average constrainers for the parameters $\alpha, q$. 

From the scatter plots, it is clear that there is a correlation between how constraining stars are and how much time they have spent being affected by the gravitational potential (see flight time panel). The most powerful stars in our golden sample are therefore tightly bound and have spent hundreds or thousands of Myr orbiting around the Galaxy. Unfortunately, part of these stars spend most of their time in a region where the Galactic Disc dominates the gravitational potential and while we have assumed perfect knowledge of this component, in reality this will hinder the halo reconstruction. In addition, the identification of these HVSs is difficult because of their low Galactocentric velocities.

On the other hand, we also identify a useful sample composed of average constrainers for $\alpha$ and strong constrainers for $q$. The stars in this sample are located at Galactocentric distances above $2$ kpc and since around half of them are moving along unbound trajectories, their identification is easier.  Notice in particular that in order to produce an average constraint for $\alpha$, a HVS needs an ejection velocity sufficiently high to reach Galactocentric distances equal to the scale radius $r_s$ and it is not required to be there when observed. This is not surprising since $\alpha$ is the effective parameter measured in the $M_s-r_s$ plane. Therefore, these distributions of Galactocentric velocities and positions set clear targets for observations aimed at measuring the Galactic halo with HVSs. Note also that while not shown, the results for halo B and C follow the same trends.

We point out that the stars in the average constrainer class also represent the main driver behind our simulated constraints, mainly because of their overwhelming number compared to the other classes. These stars follow orbits able to reach $r_s$, but at the time of observation their flight time is relatively short ($\sim 10$ Myr) and are expected not to have reached their first apocenter yet.
 
Regarding the feasibility of future observations another important factor to consider is the presence in the sample of stars which, by chance, follow HVS-compatible orbits. While quantifying the impact of this contamination and correcting for its effect in the inferred halo parameters is not the goal of this paper, we can still quantify its expected magnitude using simple arguments. \cite{Robin2012a} estimated the number of halo stars in the final \Gaia catalogue to be equal to $10^7$. Of these, around $10^4$ will have velocities higher than our golden sample threshold of $450$~km/s. We then assume an isotropic velocity distribution and a typical \Gaia HVS at a distance of $10$ kpc from the Sun, moving at $10^3$ km/s (corresponding to a $\sim10$ mas/yr proper motion) with $10\%$ parallax error, $1$ km/s radial velocity error, and $10$ $\mu$as/yr proper motion error \citep{Marchetti2018}. For this type of object, we obtain that the fraction of stars with a proper motion vector consistent with the radial direction is $\sim 10^{-3}$. According to this estimate, the number of halo stars polluting our sample would then be $\sim10$; close to the $\sim 100$ real HVSs that we expect in our average constrainer class. Notice however that this bound is particularly conservative since we have neglected additional properties, such as metallicity, that correlate with being a HVS.

\section{Discussion and conclusions}
\label{sec:conclusion}

Hypervelocity stars are remarkable objects. According to the leading model, they are ejected from the GC with high velocity (around {$10^3$ km/s}) and travel along orbits spanning at least tens of kiloparsecs. This allows them to probe the gravitational potential of the Milky Way where the dark matter halo is dominant. 

In this work we have developed a technique to extract information about the Galactic potential and the ejection mechanism from the observed HVS distribution in mass, velocity and position. Our method predicts the density of HVSs for a given stellar mass and phase-space position by back-propagating the observed location to the ejection point. The orbit is therefore required to cross the GC within a stellar lifetime to result in a non-zero distribution function. This is the basis of our likelihood pipeline, used to produce model constraints. To test our method we have applied it to mock HVS populations, designed to mimic what the European Space Agency's mission \Gaia will observe in the next few years. In our simulations, HVSs are propagated in three fiducial axisymmetric potentials and then used to reconstruct the dark matter components, modelled using a spheroidal NFW potential defined by a scale radius, scale mass and axis ratio ($r_s, M_s, q$). 

\begin{figure*}
	\includegraphics[height=0.3\textwidth]{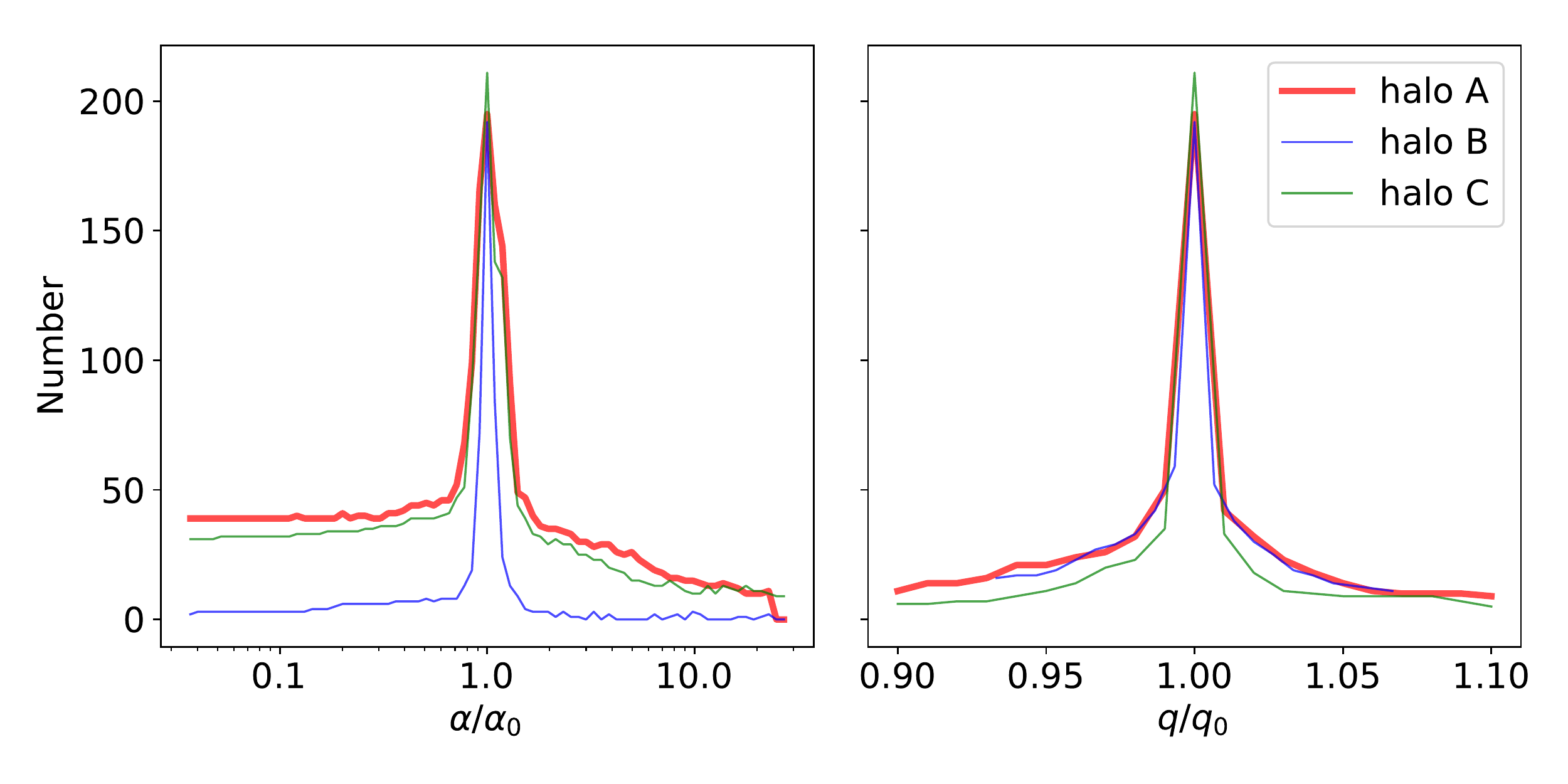}
	\includegraphics[height=0.3\textwidth]{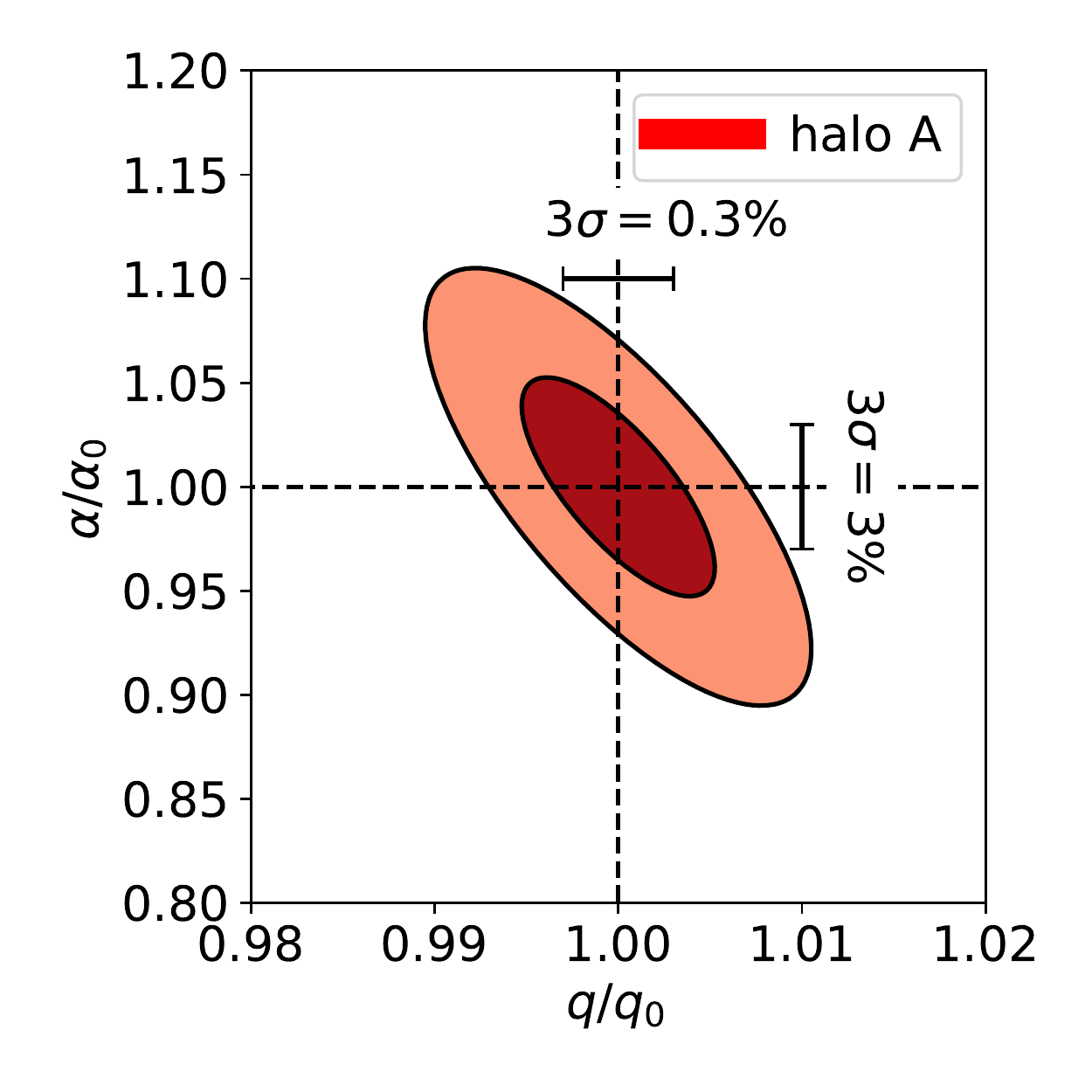}
	\caption{Summary of the simulated constraints obtained in this paper using $197$ HVSs propagated in a Galaxy with dark matter halo A. Assuming a spheroidal NFW potential, HVSs are able to measure the axis ratio $q$ and the effective scale parameter $\alpha=M_s/r_s^2$. The plots on the left show number of stars allowing a certain value for one of the parameters ($q$ or $\alpha$) when freezing the second one to its fiducial expectation ($q_0$ for one and $\alpha_0$ for the other). By looking at the likelihood evaluated along to the two directions marked by the dashed lines in the $\alpha-q$ plane, and taking into account covariance, we are able to estimate the marginalized $3\sigma$ error bars visible on the top-most and right-most side of the contour plot (see Sec.~\ref{sec:L1} and \ref{sec:L2}). An estimate for the correlation between $q$ and $\alpha$ is found in Sec.~\ref{sec:corr}. The plotted ellipses represent the $5$ and $10$ sigma contours for the inferred bivariate distribution. The figure shows the robustness of our result to changes in the fiducial values for $\alpha$ and $q$ (halo A, B, C, see Table~\ref{tab:pot}) and it illustrates how the width of the peaks in the histograms translates, through the combination of multiple stars, in tight constraints for the halo parameters.}
	\label{fig:result}
\end{figure*}

The results of our analysis are very promising, we find that $\sim 200$ HVSs are able to provide an unbiased measurement of the NFW potential parameters with sub-percent uncertainties, thanks mainly to the strict constraints we impose on the ejection location and angular momentum at that instant. While promising, it should be kept in mind that our results were obtained in an idealized scenario. We assumed perfect knowledge of the baryonic potential and the parametric form of dark matter halo, not accounting for modelling errors or {\it Gaia}-like observational uncertainties. We want to stress that while this work is interested in providing a method to measure the Galactic halo with HVSs, the technique we have developed can also be used to constrain any component of the Galactic potential or property of the HVS population. In particular, our method can be trivially generalized to constrain parametric forms of the ejection conditions like eq.~\ref{eq:Rvm}. We plan to explore this wider parameter space and the impact of observational uncertainties on the full reconstruction in a subsequent paper (Evans et al. in prep). 

We test the robustness of our results for both spherical and oblate geometries, and for two different values of the fiducial scale parameters. In all cases we observe a natural degeneracy, whereby only the combination $\alpha = M_s/r_s^2$ can be constrained. We also identify two special classes of stars, named "poor" and "strong" constrainers. The first class contains $15-30\%$ of our sample and the stars in it are not be able to produce likelihood contours because, of the model explored, only the fiducial one produces a non-zero likelihood. The second class, similarly sized, contains stars which are unable to tell the majority of the potentials in our grid from each other. If we neglect the poor constrainers, we identify a useful sample of $\sim150$ stars which can individually measure $\alpha$ with a precision of $\sim20\%$ and $q$ with a precision of $\sim 5\%$. 

We point out, however, that the full constraints depend on the sample we consider. HVSs belonging to the high-velocity tail of the ejection distribution provide the bulk of the information, but because their distribution is known to be particularly sensitive to the semi-major axis and mass distributions of binaries in the GC \citep{Rossi2013}, different models for these quantities will correspond to different constraints. 

In Fig.~\ref{fig:result} we summarize the final results of this paper by showing how many stars in our sample we expect to allow a given halo potential. It should be noted that the correlation visible in the rightmost panel of this figure is not a characteristic of individual HVSs, but arises because single HVSs constrain different combinations of $q$ and $\alpha$. To account for this in our estimates, in Sec.~\ref{sec:corr} we have estimated the covariance between the two variables in the stacked likelihood. Our final estimate of the $1\sigma$ relative uncertainties is $<1\%$ for $\alpha$ and $<0.1\%$ for $q$ (see Table~\ref{tab:classes}).

We also show that the constraining power of HVSs correlates with some observational quantities. In particular, we identify two essential properties characterizing a useful sample of HVSs: their orbits should be able to reach the NFW scale radius and their flight-time should be as long as possible. This roughly translates into distances from the GC between $2$ and $20$ kpc and Galactocentric velocities $\lesssim 900$ km/s. Furthermore, we also show that the number of contaminants moving along quasi-radial orbits by chance should be negligible.

At last, while a detailed comparison between our forecast and actual measurements of the Milky Way halo using other probes is not straightforward, we find it useful to report the result for our primary model (halo A) in a standard format. Notice that even if we assume a spherical halo, the degeneracy in the $M_s-r_s$ plane does not allow us to constrain the virial mass $M_{200}$ or the virial radius $R_{200}$.\footnote{We define $M_{200}$ as the mass inside a sphere surrounding the halo where the average density is $200$ times the critical density of the Universe at the present time. The radius of this sphere is known as virial radius, $R_{200}$.} This degeneracy can be broken if we assume that the Milky Way concentration parameter $c=R_{200}/r_s$ is related to the virial mass through a mass-concentration relation, as seen in $\Lambda$CDM numerical simulations \citep[see, e.g.][]{Navarro1997}. Without assuming a spherical halo, we use the relation from \cite{Dutton2014} and the latest Planck 2015 cosmology \cite{PlanckCollaboration2015} to translate our precision in $\alpha$ into a virial mass of $\log_{10}
(M_{200}/M_\odot) = (12.14\pm 0.02)$, corresponding to $\sim10\%$ precision in $M_{200}$.  Notice that, although our reconstruction of $\alpha$ is not affected by bias, the recovered virial mass contours do not include the true $M_{200} = 1.04\times 10^{12}$ M$_\odot$ corresponding to the fiducial halo A. As observed before in \cite{Wang2015}, this is a perfect example of how assumptions, like imposing a mass-concentration relation, can affect the results obtained with dynamical tracers. A direct comparison of our forecast with the constraints of other probes provided by the same paper (their figure~1) also suggests that our technique is able to achieve competitive results. 

Our conclusions paint an optimistic picture for the introduction of HVSs as a new dynamical tracer of the Galactic potential, especially when combined with the prospects of HVS detections in the final release of \Gaia \citep{Marchetti2018}. The wealth of data that will become available in the next few years will allow measurements of the dark matter distribution in the Milky Way of unprecedented precision. However, in order to produce accurate results and combine the information provided by multiple tracers, particular care should be taken and modelling biases be carefully considered. 

\section*{Acknowledgements}
We thank the referee for the useful comments regarding the manuscript. We also thank Re'em Sari and Yuri Levin for useful discussion.  OC is supported by a de Sitter Fellowship of the Netherlands Organization for Scientific Research (NWO). TM and EMR acknowledge support from NWO TOP grant Module 2, project number 614.001.401.



\bibliographystyle{mnras}
\bibliography{HVS_forecast} 

\begin{thebibliography}{}
\makeatletter
\relax
\def\mn@urlcharsother{\let\do\@makeother \do\$\do\&\do\#\do\^\do\_\do\%\do\~}
\def\mn@doi{\begingroup\mn@urlcharsother \@ifnextchar [ {\mn@doi@}
  {\mn@doi@[]}}
\def\mn@doi@[#1]#2{\def\@tempa{#1}\ifx\@tempa\@empty \href
  {http://dx.doi.org/#2} {doi:#2}\else \href {http://dx.doi.org/#2} {#1}\fi
  \endgroup}
\def\mn@eprint#1#2{\mn@eprint@#1:#2::\@nil}
\def\mn@eprint@arXiv#1{\href {http://arxiv.org/abs/#1} {{\tt arXiv:#1}}}
\def\mn@eprint@dblp#1{\href {http://dblp.uni-trier.de/rec/bibtex/#1.xml}
  {dblp:#1}}
\def\mn@eprint@#1:#2:#3:#4\@nil{\def\@tempa {#1}\def\@tempb {#2}\def\@tempc
  {#3}\ifx \@tempc \@empty \let \@tempc \@tempb \let \@tempb \@tempa \fi \ifx
  \@tempb \@empty \def\@tempb {arXiv}\fi \@ifundefined
  {mn@eprint@\@tempb}{\@tempb:\@tempc}{\expandafter \expandafter \csname
  mn@eprint@\@tempb\endcsname \expandafter{\@tempc}}}

\bibitem[\protect\citeauthoryear{Battaglia et~al.,}{Battaglia
  et~al.}{2005}]{Battaglia2005}
Battaglia G.,  et~al., 2005, \mn@doi [Monthly Notices of the Royal Astronomical
  Society] {10.1111/j.1365-2966.2005.09367.x}, 364, 433

\bibitem[\protect\citeauthoryear{Bovy}{Bovy}{2015}]{Bovy2015a}
Bovy J.,  2015, \mn@doi [The Astrophysical Journal Supplement Series]
  {10.1088/0067-0049/216/2/29}, 216, 29

\bibitem[\protect\citeauthoryear{Bovy, Bird, P{\'{e}}rez, Majewski, Nidever  \&
  Zasowski}{Bovy et~al.}{2015}]{Bovy2015}
Bovy J.,  Bird J.~C.,  P{\'{e}}rez A. E.~G.,  Majewski S.~R.,  Nidever D.~L.,
  Zasowski G.,  2015, \mn@doi [The Astrophysical Journal]
  {10.1088/0004-637X/800/2/83}, 800, 83

\bibitem[\protect\citeauthoryear{Bovy, Rix, Green, Schlafly  \&
  Finkbeiner}{Bovy et~al.}{2016a}]{Bovy2016a}
Bovy J.,  Rix H.-W.,  Green G.~M.,  Schlafly E.~F.,   Finkbeiner D.~P.,  2016a,
  \mn@doi [The Astrophysical Journal] {10.3847/0004-637X/818/2/130}, 818, 130

\bibitem[\protect\citeauthoryear{Bovy, Bahmanyar, Fritz  \& Kallivayalil}{Bovy
  et~al.}{2016b}]{Bovy2016}
Bovy J.,  Bahmanyar A.,  Fritz T.~K.,   Kallivayalil N.,  2016b, \mn@doi [The
  Astrophysical Journal] {10.3847/1538-4357/833/1/31}, 833, 31

\bibitem[\protect\citeauthoryear{Bowden, Evans  \& Williams}{Bowden
  et~al.}{2016}]{Bowden2016}
Bowden A.,  Evans N.~W.,   Williams A.~A.,  2016, \mn@doi [Monthly Notices of
  the Royal Astronomical Society] {10.1093/mnras/stw994}, 460, 329

\bibitem[\protect\citeauthoryear{Brown}{Brown}{2015}]{Brown2015}
Brown W.~R.,  2015, \mn@doi [Annual Review of Astronomy and Astrophysics]
  {10.1146/annurev-astro-082214-122230}, 53, 15

\bibitem[\protect\citeauthoryear{Brown, Geller, Kenyon  \& Kurtz}{Brown
  et~al.}{2005}]{Brown2005}
Brown W.~R.,  Geller M.~J.,  Kenyon S.~J.,   Kurtz M.~J.,  2005, \mn@doi [The
  Astrophysical Journal] {10.1086/429378}, 622, L33

\bibitem[\protect\citeauthoryear{{Brown}, {Geller}, {Kenyon}, {Kurtz}  \&
  {Bromley}}{{Brown} et~al.}{2007}]{Brown2007}
{Brown} W.~R.,  {Geller} M.~J.,  {Kenyon} S.~J.,  {Kurtz} M.~J.,   {Bromley}
  B.~C.,  2007, \mn@doi [\apj] {10.1086/513595}, \href
  {https://ui.adsabs.harvard.edu/\#abs/2007ApJ...660..311B} {660, 311}

\bibitem[\protect\citeauthoryear{Brown, Geller, Kenyon  \& Diaferio}{Brown
  et~al.}{2010}]{Brown2010}
Brown W.~R.,  Geller M.~J.,  Kenyon S.~J.,   Diaferio A.,  2010, \mn@doi [The
  Astronomical Journal] {10.1088/0004-6256/139/1/59}, 139, 59

\bibitem[\protect\citeauthoryear{Brown, Geller  \& Kenyon}{Brown
  et~al.}{2014}]{Brown2014}
Brown W.~R.,  Geller M.~J.,   Kenyon S.~J.,  2014, \mn@doi [The Astrophysical
  Journal] {10.1088/0004-637X/787/1/89}, 787, 89

\bibitem[\protect\citeauthoryear{Brown, Lattanzi, Kenyon  \& Geller}{Brown
  et~al.}{2018}]{Brown2018}
Brown W.~R.,  Lattanzi M.~G.,  Kenyon S.~J.,   Geller M.~J.,  2018

\bibitem[\protect\citeauthoryear{Cacciari, Pancino  \& Bellazzini}{Cacciari
  et~al.}{2016}]{Cacciari2016}
Cacciari C.,  Pancino E.,   Bellazzini M.,  2016, \mn@doi [Astronomische
  Nachrichten] {10.1002/asna.201612394}, 337, 899

\bibitem[\protect\citeauthoryear{Debattista, Moore, Quinn, Kazantzidis, Maas,
  Mayer, Read  \& Stadel}{Debattista et~al.}{2008}]{Debattista2008}
Debattista V.~P.,  Moore B.,  Quinn T.,  Kazantzidis S.,  Maas R.,  Mayer L.,
  Read J.,   Stadel J.,  2008, \mn@doi [The Astrophysical Journal]
  {10.1086/587977}, 681, 1076

\bibitem[\protect\citeauthoryear{Deg \& Widrow}{Deg \& Widrow}{2013}]{Deg2013}
Deg N.,  Widrow L.,  2013, \mn@doi [Monthly Notices of the Royal Astronomical
  Society] {10.1093/mnras/sts089}, 428, 912

\bibitem[\protect\citeauthoryear{Drimmel, Cabrera-Lavers  \&
  L{\'{o}}pez-Corredoira}{Drimmel et~al.}{2003}]{Drimmel2003}
Drimmel R.,  Cabrera-Lavers A.,   L{\'{o}}pez-Corredoira M.,  2003, \mn@doi
  [Astronomy {\&} Astrophysics] {10.1051/0004-6361:20031070}, 409, 205

\bibitem[\protect\citeauthoryear{Duffau, {Katherina Vivas}, Zinn, M{\'{e}}ndez
  \& Ruiz}{Duffau et~al.}{2014}]{Duffau2014}
Duffau S.,  {Katherina Vivas} A.,  Zinn R.,  M{\'{e}}ndez R.~A.,   Ruiz M.~T.,
  2014, \mn@doi [Astronomy {\&} Astrophysics] {10.1051/0004-6361/201219654},
  566, A118

\bibitem[\protect\citeauthoryear{Dutton \& Macci{\`{o}}}{Dutton \&
  Macci{\`{o}}}{2014}]{Dutton2014}
Dutton A.~A.,  Macci{\`{o}} A.~V.,  2014, \mn@doi [Monthly Notices of the Royal
  Astronomical Society] {10.1093/mnras/stu742}, 441, 3359

\bibitem[\protect\citeauthoryear{Eisenhauer et~al.,}{Eisenhauer
  et~al.}{2005}]{Eisenhauer2005}
Eisenhauer F.,  et~al., 2005, \mn@doi [The Astrophysical Journal]
  {10.1086/430667}, 628, 246

\bibitem[\protect\citeauthoryear{Foreman-Mackey, Hogg, Lang  \&
  Goodman}{Foreman-Mackey et~al.}{2013}]{Foreman-Mackey2013}
Foreman-Mackey D.,  Hogg D.~W.,  Lang D.,   Goodman J.,  2013, \mn@doi
  [Publications of the Astronomical Society of the Pacific] {10.1086/670067},
  125, 306

\bibitem[\protect\citeauthoryear{Fragione \& Loeb}{Fragione \&
  Loeb}{2017}]{Fragione2017}
Fragione G.,  Loeb A.,  2017, \mn@doi [New Astronomy]
  {10.1016/J.NEWAST.2017.03.002}, 55, 32

\bibitem[\protect\citeauthoryear{{Gaia Collaboration}}{{Gaia
  Collaboration}}{2016}]{GaiaCollaboration2016}
{Gaia Collaboration} 2016, \mn@doi [Astronomy {\&} Astrophysics]
  {10.1051/0004-6361/201629272}, 595, A1

\bibitem[\protect\citeauthoryear{{Gaia Collaboration}, Brown, Vallenari,
  Prusti, de Bruijne, Babusiaux  \& Bailer-Jones}{{Gaia Collaboration}
  et~al.}{2018}]{GaiaCollaboration2018}
{Gaia Collaboration} G.,  Brown A. G.~A.,  Vallenari A.,  Prusti T.,  de
  Bruijne J. H.~J.,  Babusiaux C.,   Bailer-Jones C. A.~L.,  2018

\bibitem[\protect\citeauthoryear{Garrett \& Duda}{Garrett \&
  Duda}{2011}]{Garrett2011}
Garrett K.,  Duda G.,  2011, \mn@doi [Advances in Astronomy]
  {10.1155/2011/968283}, 2011, 1

\bibitem[\protect\citeauthoryear{Genzel, Eisenhauer  \& Gillessen}{Genzel
  et~al.}{2010}]{Genzel2010}
Genzel R.,  Eisenhauer F.,   Gillessen S.,  2010, \mn@doi [Reviews of Modern
  Physics] {10.1103/RevModPhys.82.3121}, 82

\bibitem[\protect\citeauthoryear{Ghez et~al.,}{Ghez et~al.}{2003}]{Ghez2003}
Ghez A.~M.,  et~al., 2003, \mn@doi [The Astrophysical Journal]
  {10.1086/374804}, 586, L127

\bibitem[\protect\citeauthoryear{Ghez et~al.,}{Ghez et~al.}{2008}]{Ghez2008}
Ghez A.~M.,  et~al., 2008, \mn@doi [The Astrophysical Journal]
  {10.1086/592738}, 689, 1044

\bibitem[\protect\citeauthoryear{Gibbons, Belokurov  \& Evans}{Gibbons
  et~al.}{2014}]{Gibbons2014}
Gibbons S. L.~J.,  Belokurov V.,   Evans N.~W.,  2014, \mn@doi [Monthly Notices
  of the Royal Astronomical Society] {10.1093/mnras/stu1986}, 445, 3788

\bibitem[\protect\citeauthoryear{Gnedin, Gould, Miralda‐Escude  \&
  Zentner}{Gnedin et~al.}{2005}]{Gnedin2005}
Gnedin O.~Y.,  Gould A.,  Miralda‐Escude J.,   Zentner A.~R.,  2005, \mn@doi
  [The Astrophysical Journal] {10.1086/496958}, 634, 344

\bibitem[\protect\citeauthoryear{Goodman \& Weare}{Goodman \&
  Weare}{2010}]{Goodman2010}
Goodman J.,  Weare J.,  2010, \mn@doi [Communications in Applied Mathematics
  and Computational Science] {10.2140/camcos.2010.5.65}, 5, 65

\bibitem[\protect\citeauthoryear{Green et~al.,}{Green
  et~al.}{2015}]{Green2015a}
Green G.~M.,  et~al., 2015, \mn@doi [The Astrophysical Journal]
  {10.1088/0004-637X/810/1/25}, 810, 25

\bibitem[\protect\citeauthoryear{Hernquist}{Hernquist}{1990}]{Hernquist1990}
Hernquist L.,  1990, \mn@doi [The Astrophysical Journal] {10.1086/168845}, 356,
  359

\bibitem[\protect\citeauthoryear{Hills}{Hills}{1988}]{Hills1988}
Hills J.~G.,  1988, \mn@doi [Nature] {10.1038/331687a0}, 331, 687

\bibitem[\protect\citeauthoryear{Hoekstra, Bartelmann, Dahle, Israel, Limousin
  \& Meneghetti}{Hoekstra et~al.}{2013}]{Hoekstra2013}
Hoekstra H.,  Bartelmann M.,  Dahle H.,  Israel H.,  Limousin M.,   Meneghetti
  M.,  2013, \mn@doi [Space Science Reviews] {10.1007/s11214-013-9978-5}, 177,
  75

\bibitem[\protect\citeauthoryear{Huang et~al.,}{Huang et~al.}{2016}]{Huang2016}
Huang Y.,  et~al., 2016, \mn@doi [Monthly Notices of the Royal Astronomical
  Society] {10.1093/mnras/stw2096}, 463, 2623

\bibitem[\protect\citeauthoryear{Hurley, Pols  \& Tout}{Hurley
  et~al.}{2000}]{Hurley2000a}
Hurley J.~R.,  Pols O.~R.,   Tout C.~A.,  2000, \mn@doi [Monthly Notices of the
  Royal Astronomical Society] {10.1046/j.1365-8711.2000.03426.x}, 315, 543

\bibitem[\protect\citeauthoryear{Johnston, Spergel  \& Hernquist}{Johnston
  et~al.}{1995}]{Johnston1995}
Johnston K.~V.,  Spergel D.~N.,   Hernquist L.,  1995, \mn@doi [The
  Astrophysical Journal] {10.1086/176247}, 451, 598

\bibitem[\protect\citeauthoryear{Jordi et~al.,}{Jordi et~al.}{2010}]{Jordi2010}
Jordi C.,  et~al., 2010, \mn@doi [Astronomy {\&} Astrophysics]
  {10.1051/0004-6361/201015441}, 523, A48

\bibitem[\protect\citeauthoryear{Katz et~al.,}{Katz et~al.}{2018}]{Katz2018}
Katz D.,  et~al., 2018

\bibitem[\protect\citeauthoryear{{King III}, Brown, Geller  \& Kenyon}{{King
  III} et~al.}{2015}]{III2015}
{King III} C.,  Brown W.~R.,  Geller M.~J.,   Kenyon S.~J.,  2015, \mn@doi [The
  Astrophysical Journal] {10.1088/0004-637X/813/2/89}, 813, 89

\bibitem[\protect\citeauthoryear{Klypin, Kravtsov, Valenzuela  \& Prada}{Klypin
  et~al.}{1999}]{Klypin1999}
Klypin A.,  Kravtsov A.~V.,  Valenzuela O.,   Prada F.,  1999, \mn@doi [The
  Astrophysical Journal] {10.1086/307643}, 522, 82

\bibitem[\protect\citeauthoryear{Law \& Majewski}{Law \&
  Majewski}{2010}]{Law2010}
Law D.~R.,  Majewski S.~R.,  2010, \mn@doi [The Astrophysical Journal]
  {10.1088/0004-637X/714/1/229}, 714, 229

\bibitem[\protect\citeauthoryear{Law, Majewski  \& Johnston}{Law
  et~al.}{2009}]{Law2009}
Law D.~R.,  Majewski S.~R.,   Johnston K.~V.,  2009, \mn@doi [The Astrophysical
  Journal] {10.1088/0004-637X/703/1/L67}, 703, L67

\bibitem[\protect\citeauthoryear{Loebman et~al.,}{Loebman
  et~al.}{2014}]{Loebman2014}
Loebman S.~R.,  et~al., 2014, \mn@doi [The Astrophysical Journal]
  {10.1088/0004-637X/794/2/151}, 794, 151

\bibitem[\protect\citeauthoryear{Mandelbaum}{Mandelbaum}{2014}]{Mandelbaum2014}
Mandelbaum R.,  2014, \mn@doi [Proceedings of the International Astronomical
  Union] {10.1017/S1743921315003452}, 10, 86

\bibitem[\protect\citeauthoryear{Marchetti, Contigiani, Rossi, Albert, Brown
  \& Sesana}{Marchetti et~al.}{2018}]{Marchetti2018}
Marchetti T.,  Contigiani O.,  Rossi E.~M.,  Albert J.~G.,  Brown A. G.~A.,
  Sesana A.,  2018, \mn@doi [Monthly Notices of the Royal Astronomical Society]
  {10.1093/mnras/sty579}, 476, 4697

\bibitem[\protect\citeauthoryear{Marshall, Robin, Reyl{\'{e}}, Schultheis  \&
  Picaud}{Marshall et~al.}{2006}]{Marshall2006}
Marshall D.~J.,  Robin A.~C.,  Reyl{\'{e}} C.,  Schultheis M.,   Picaud S.,
  2006, \mn@doi [Astronomy {\&} Astrophysics] {10.1051/0004-6361:20053842},
  453, 635

\bibitem[\protect\citeauthoryear{Miyamoto \& Nagai}{Miyamoto \&
  Nagai}{1975}]{Miyamoto1975}
Miyamoto M.,  Nagai R.,  1975, Publications of the Astronomical Society of
  Japan, 27, 533

\bibitem[\protect\citeauthoryear{Moore, Ghigna, Governato, Lake, Quinn, Stadel
  \& Tozzi}{Moore et~al.}{1999}]{Moore1999}
Moore B.,  Ghigna S.,  Governato F.,  Lake G.,  Quinn T.,  Stadel J.,   Tozzi
  P.,  1999, \mn@doi [The Astrophysical Journal] {10.1086/312287}, 524, L19

\bibitem[\protect\citeauthoryear{Navarro, Frenk  \& White}{Navarro
  et~al.}{1997}]{Navarro1997}
Navarro J.~F.,  Frenk C.~S.,   White S. D.~M.,  1997, \mn@doi [The
  Astrophysical Journal] {10.1086/304888}, 490, 493

\bibitem[\protect\citeauthoryear{Newberg et~al.,}{Newberg
  et~al.}{2002}]{Newberg2002}
Newberg H.~J.,  et~al., 2002, \mn@doi [The Astrophysical Journal]
  {10.1086/338983}, 569, 245

\bibitem[\protect\citeauthoryear{Odenkirchen et~al.,}{Odenkirchen
  et~al.}{2001}]{Odenkirchen2001}
Odenkirchen M.,  et~al., 2001, \mn@doi [The Astrophysical Journal]
  {10.1086/319095}, 548, L165

\bibitem[\protect\citeauthoryear{Pearson, K{\"{u}}pper, Johnston  \&
  Price-Whelan}{Pearson et~al.}{2015}]{Pearson2015}
Pearson S.,  K{\"{u}}pper A. H.~W.,  Johnston K.~V.,   Price-Whelan A.~M.,
  2015, \mn@doi [The Astrophysical Journal] {10.1088/0004-637X/799/1/28}, 799,
  28

\bibitem[\protect\citeauthoryear{Perets, Wu, Zhao, Famaey, Gentile  \&
  Alexander}{Perets et~al.}{2009}]{Perets2009}
Perets H.~B.,  Wu X.,  Zhao H.,  Famaey B.,  Gentile G.,   Alexander T.,  2009,
  \mn@doi [The Astrophysical Journal] {10.1088/0004-637X/697/2/2096}, 697, 2096

\bibitem[\protect\citeauthoryear{Peter, Rocha, Bullock  \& Kaplinghat}{Peter
  et~al.}{2013}]{Peter2013}
Peter A. H.~G.,  Rocha M.,  Bullock J.~S.,   Kaplinghat M.,  2013, \mn@doi
  [Monthly Notices of the Royal Astronomical Society] {10.1093/mnras/sts535},
  430, 105

\bibitem[\protect\citeauthoryear{{Planck Collaboration}}{{Planck
  Collaboration}}{2016}]{PlanckCollaboration2015}
{Planck Collaboration} 2016, \mn@doi [Astronomy {\&} Astrophysics]
  {10.1051/0004-6361/201525830}, 594, A13

\bibitem[\protect\citeauthoryear{Portail, Gerhard, Wegg  \& Ness}{Portail
  et~al.}{2017}]{Portail2017}
Portail M.,  Gerhard O.,  Wegg C.,   Ness M.,  2017, \mn@doi [Monthly Notices
  of the Royal Astronomical Society] {10.1093/mnras/stw2819}, 465, 1621

\bibitem[\protect\citeauthoryear{Posti \& Helmi}{Posti \&
  Helmi}{2018}]{Posti2018}
Posti L.,  Helmi A.,  2018

\bibitem[\protect\citeauthoryear{Price-Whelan, Hogg, Johnston  \&
  Hendel}{Price-Whelan et~al.}{2014}]{Price-Whelan2014}
Price-Whelan A.~M.,  Hogg D.~W.,  Johnston K.~V.,   Hendel D.,  2014, \mn@doi
  [The Astrophysical Journal] {10.1088/0004-637X/794/1/4}, 794, 4

\bibitem[\protect\citeauthoryear{Reid \& Dame}{Reid \& Dame}{2016}]{Reid2016}
Reid M.~J.,  Dame T.~M.,  2016, \mn@doi [The Astrophysical Journal]
  {10.3847/0004-637X/832/2/159}, 832, 159

\bibitem[\protect\citeauthoryear{Robin et~al.,}{Robin
  et~al.}{2012}]{Robin2012a}
Robin A.~C.,  et~al., 2012, \mn@doi [Astronomy {\&} Astrophysics]
  {10.1051/0004-6361/201118646}, 543, A100

\bibitem[\protect\citeauthoryear{Rossi, Kobayashi  \& Sari}{Rossi
  et~al.}{2014}]{Rossi2013}
Rossi E.~M.,  Kobayashi S.,   Sari R.,  2014, \mn@doi [The Astrophysical
  Journal] {10.1088/0004-637X/795/2/125}, 795, 125

\bibitem[\protect\citeauthoryear{Rossi, Marchetti, Cacciato, Kuiack  \&
  Sari}{Rossi et~al.}{2017}]{Rossi2016}
Rossi E.~M.,  Marchetti T.,  Cacciato M.,  Kuiack M.,   Sari R.,  2017, \mn@doi
  [Monthly Notices of the Royal Astronomical Society] {10.1093/mnras/stx098},
  12, stx098

\bibitem[\protect\citeauthoryear{Sari, Kobayashi  \& Rossi}{Sari
  et~al.}{2010}]{Sari2010}
Sari R.,  Kobayashi S.,   Rossi E.~M.,  2010, \mn@doi [The Astrophysical
  Journal] {10.1088/0004-637X/708/1/605}, 708, 605

\bibitem[\protect\citeauthoryear{Sesana, Haardt  \& Madau}{Sesana
  et~al.}{2007}]{Sesana2007}
Sesana A.,  Haardt F.,   Madau P.,  2007, \mn@doi [Monthly Notices of the Royal
  Astronomical Society: Letters] {10.1111/j.1745-3933.2007.00331.x}, 379, L45

\bibitem[\protect\citeauthoryear{Vera-Ciro \& Helmi}{Vera-Ciro \&
  Helmi}{2013}]{Vera-Ciro2013}
Vera-Ciro C.,  Helmi A.,  2013, ] {10.1088/2041-8205/773/1/L4}, 4, 2

\bibitem[\protect\citeauthoryear{Wang et~al.,}{Wang et~al.}{2011}]{Wang2011}
Wang J.,  et~al., 2011, \mn@doi [Monthly Notices of the Royal Astronomical
  Society] {10.1111/j.1365-2966.2011.18220.x}, 413, 1373

\bibitem[\protect\citeauthoryear{Wang, Han, Cooper, Cole, Frenk  \&
  Lowing}{Wang et~al.}{2015}]{Wang2015}
Wang W.,  Han J.,  Cooper A.~P.,  Cole S.,  Frenk C.,   Lowing B.,  2015,
  \mn@doi [Monthly Notices of the Royal Astronomical Society]
  {10.1093/mnras/stv1647}, 453, 377

\bibitem[\protect\citeauthoryear{Westera, Lejeune, Buser, Cuisinier  \&
  Bruzual}{Westera et~al.}{2002}]{Westera2002}
Westera P.,  Lejeune T.,  Buser R.,  Cuisinier F.,   Bruzual G.,  2002, \mn@doi
  [Astronomy {\&} Astrophysics] {10.1051/0004-6361:20011493}, 381, 524

\bibitem[\protect\citeauthoryear{Yu \& Madau}{Yu \& Madau}{2007}]{Yu2007}
Yu Q.,  Madau P.,  2007, \mn@doi [Monthly Notices of the Royal Astronomical
  Society] {10.1111/j.1365-2966.2007.12034.x}, 379, 1293

\bibitem[\protect\citeauthoryear{Yu \& Tremaine}{Yu \& Tremaine}{2003}]{Yu2003}
Yu Q.,  Tremaine S.,  2003, \mn@doi [The Astrophysical Journal]
  {10.1086/379546}, 599, 1129

\makeatother
\end{thebibliography}



\appendix

\section{Two-body encounters}
\label{sec:stability}

In the main section of the paper, we have not taken into account the deflection of HVS orbits due to two-body encounters with the Galactic stellar distribution. In this Appendix, we verify that these perturbations no dot affect our conclusions significantly. We consider here the motion of a star representative of the average constrainer class in our mock catalogues with velocity $v=1000$ km/s and mass $m = 1$ M$_\odot$ moving along a quasi-radial orbit. For simplicity we also assume the stellar component of the Galaxy to be composed of $1$ M$_\odot$ objects.

The close encounter between two objects of mass $m$ with impact parameter $b$ and relative velocity $v$ is expected to induce a velocity kick of the order of $\delta v= 2Gm/(b v)$. If we model the galaxy as a collisionless system with number density $n$, then the average time between two encounters such that $\delta v > v$ is
\begin{equation}
t_\mathrm{CE} = \frac{v^3}{4 \pi G^2 m^2 n}.
\end{equation}
Even in the dense environment of the galactic centre with $n= 10^6$ pc$^{-3}$ we obtain $t_\mathrm{CE}$ two orders of magnitude larger than the age of the Universe. Therefore, strong encounters are not expected to influence the HVS orbits considered in our main sections.

Despite this, encounters with larger impact parameters will still have an effect. The size of velocity deflection experienced in a time $t_{\mathrm{cross}}$ can be estimated by integrating over every possible impact parameter. This is found to be equal to:
\begin{equation}
(\Delta v)^2  = \frac{8\pi G^2  m^2 \ln \Lambda}{v} t_{\mathrm{cross}},
\end{equation}
where $\ln \Lambda$ is the Coulomb logarithm equal to $\sim21$ for the typical size of the Galactic bulge $r_\mathrm{bulge} = t_\mathrm{cross} v = 1$ kpc. We focus on the bulge because it is a high-density stellar region representing the dominant source of perturbation. Assuming a typical bulge number density of $10$ pc$^{-3}$ we obtain $\Delta v\sim 10^{-2}$ km/s.

While this change might appear small, we stress that our fitting technique is based on integrating the observed HVS position back in time and imposing extremely stringent conditions on the ejection conditions. We can translate the $\Delta v$ we obtained above into ejection conditions using some simplifying assumptions. If we neglect the presence of a Galactic disk and assume radial orbits, angular momentum is conserved and is expected not to be zero, but of the order $\delta L =  r_\mathrm{bulge} \delta v \sim 10$ {km $\times$ pc/s}. Notice that this value is equal to the smoothing parameter $\sigma_L$ we have presented in Sec.~\ref{sec:methods} to impose an artificially enlarged ejection region. 
\section{Triaxial halo}
\label{sec:triaxial}

\begin{figure}
	\includegraphics[width=0.48\textwidth]{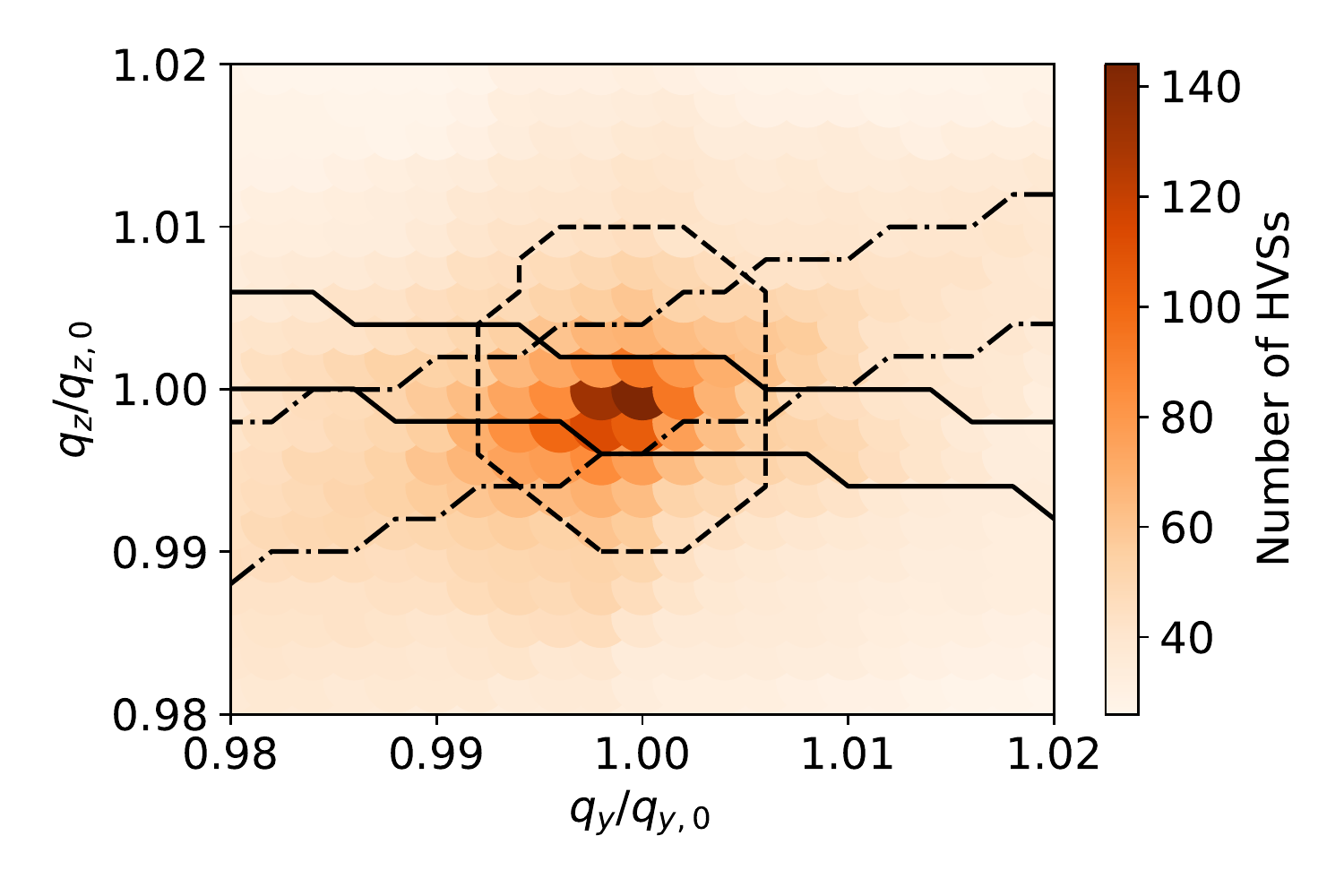}
	\caption{Number of HVSs propagated in the triaxial halo described in this Appendix for which a given halo, parametrized by the shape parameters $q_y, q_z$, is allowed.  The peak at $(1, 1)$ marks the fiducial values for the triaxial halo presented in this section ($q_{y, 0} = 0.75$ and $q_{z, 0} = 1.5$). The figure was created using stars for which a visible spread in the likelihood is present in our grid. For illustrative purposes, the dashed, dot-dashed and solid lines delimit the regions allowed by three individual stars.}
	\label{fig:qq}
\end{figure}

In this Appendix, we discuss the possibility of constraining a triaxial shape of the dark matter halo using the same method developed in the main section. Because we found that the shape of a spheroidal halo is heavily constrained, we study here if such precision can be generalized to a more complex configuration.

We extend the Galactic potential model introduced in 
Sec.~\ref{sec:pot} by changing the spheroidal NFW profile to an ellipsoidal distribution:
\begin{align}
\rho_{\text{NFW}}(x, y, z) = \frac{M_s}{4\pi r_h^3}\frac{1}{(\xi/r_s)(1+\xi/r_s)^2}, && \xi^2 = x^2 + \frac{y^2}{q_y^2} + \frac{z^2}{q_z^2},
\label{eq:triax}
\end{align}
where $q_z$ and $q_y$ define the axis ratios of the ellipsoid in the $z$ and $y$ directions with respect to the $x$ direction. For these two parameters we chose fiducial values $q_{z, 0} = 1.5$ and $ q_{y, 0} = 0.75$, while for the scale parameters $M_s$ and $r_s$ we chose the fiducial values $0.76\times10^{12}$ M$_\odot$ and $24.8$ kpc from our halo A. Notice that for $q_z = q$ and $q_y=1$ the model in eq.~\ref{eq:triax} reduces to the one used in the main text. 

Using the procedure described in Sec.~\ref{sec:mock} we build a mock catalogue of HVSs inside this Galactic potential and produce a sample of $199$ stars within the \Gaia horizon. Using the methods described in Sec.~\ref{sec:res} we then explore the plane $M_s-r_s$ using these same stars to quantify how precisely these objects can be used to constrain the Galactic dark matter halo. 

We find that, for a triaxial halo, the distribution of HVSs in the three categories of average, strong and poor constrainers is significantly different from the cases explored in the main text. For this halo, almost half of the sample ($94$ out of $199$ stars) belongs to the strong constrainer category and only $2$ stars are considered poor constrainers. This suggests that in a triaxial halo almost every HVS can provide some information about the halo. The average amount of information per star is however lower and their combination achieves a precision $\hat{\sigma}_\alpha/\hat{\sigma}=0.49$\%. This number should be compared to the lower limits on $\hat{\sigma}_\alpha$ reported in Table~\ref{tab:classes} since it does not take into account the covariance with the shape parameters. In all three cases, the values are within a factor $2$ of each other.

To study how the shape of a triaxial halo can be constrained by HVSs, we present the results of the exploration of the plane $q_y-q_z$ in Fig.~\ref{fig:qq}. In this figure, we show the number of HVSs with non-zero likelihood for a given choice of these two parameters. Notice that the values of $M_s, r_s$ are kept at their fiducial values when producing this distribution. Unlike the scale parameters, there is not a single combination of $q_z$ and $q_y$ which is constrained by every single star. To illustrate this, we have plotted the regions allowed by three of our stars. 

On average, the contours favour a positive degeneracy between the two parameters, where more asymmetric configurations in one direction are compensated by a rounded shape in the other. We point out, however, that because this is not a natural degeneracy in this parameter space, we expect this behaviour not to be a general prediction: the inclusion of observational errors or extensions of the parameter space will affect this result. 


\bsp	
\label{lastpage}
\end{document}